\documentclass[default]{aastex631}

\usepackage{amsmath}
\usepackage{comment}
\usepackage{CJK}
\usepackage{graphicx}
\usepackage{natbib}
\usepackage{multirow}
\usepackage{makecell}
\usepackage{mwe}

\usepackage{hyperref}

\newcommand{\TODO}[1]{}

\received{May 22, 2024}
\revised{November 20, 2024}
\accepted{\today}

\begin{document}
\begin{CJK*}{UTF8}{gbsn}

\title{Habitability in 4-D: Predicting the Climates of Earth Analogs across Rotation and Orbital Configurations}

\correspondingauthor{Arthur D. Adams}
\email{arthur@virginia.edu}
\author[0000-0002-7139-3695]{Arthur D. Adams}
\affiliation{Department of Astronomy \\
University of Virginia\\
530 McCormick Road, Charlottesville VA 22904}
\affiliation{Department of Earth and Planetary Sciences \\
University of California \\
900 University Ave., Riverside, CA 92521, USA}

\author{Christopher Colose}
\affiliation{Business Integra, NASA Goddard Institute for Space Studies, New York, NY 10025, USA}

\author[0000-0002-5138-8098]{Aronne Merrelli}
\affiliation{Department of Climate and Space Sciences and Engineering \\
University of Michigan \\
2455 Hayward St., Ann Arbor, MI 48109, USA}

\author[0000-0002-0569-1643]{Margaret Turnbull}
\affiliation{Carl Sagan Center for Research, SETI Institute}

\author[0000-0002-7084-0529]{Stephen R. Kane}
\affiliation{Department of Earth and Planetary Sciences \\
University of California \\
900 University Ave., Riverside, CA 92521, USA}

\begin{abstract}
Earth-like planets in the circumstellar habitable zone (HZ) may have dramatically different climate outcomes depending on their spin-orbit parameters, altering their habitability for life as we know it. We present a suite of 93 ROCKE-3D general circulation models (GCMs) for planets with the same surface conditions and average annual insolation as Earth, but with a wide range of rotation periods, obliquities, orbital eccentricities, and longitudes of periastra. Our habitability metric $f_\mathrm{HZ}$ is calculated based on the temperature and precipitation in each model across grid cells over land. Latin Hypercube Sampling (LHS) aids in sampling all 4 of the spin-orbit parameters with a computationally feasible number of GCM runs. Statistical emulation then allows us to model $f_\mathrm{HZ}$ as a smooth function with built-in estimates of statistical uncertainty. We fit our emulator to an initial set of 46 training runs, then test with an additional 46 runs at different spin-orbit values. Our emulator predicts the directly GCM-modeled habitability values for the test runs at the appropriate level of accuracy and precision. For orbital eccentricities up to 0.225, rotation period remains the primary driver of the fraction of land that remains above freezing and with precipitation above a threshold value. For rotation periods greater than $\sim 20$ days, habitability drops significantly (from $\sim 70$\% to $\sim 20$\%), driven primarily by cooler land temperatures. Obliquity is a significant secondary factor for rotation periods less than $\sim 20$ Earth days, with a factor of two impact on habitability that is maximized at intermediate obliquity.
\end{abstract}

\section{Introduction} \label{sec:intro}
The search for potentially habitable exoplanets is a high priority in NASA's current Astrophysics portfolio. The Nancy Grace Roman Space Telescope, launching in mid-2027, will include a Coronagraph Instrument to advance the technology needed to directly image mature planetary systems like our own. Following on this technology demonstration, the Habitable Worlds Observatory (HWO) is envisioned as a large infrared/optical/ultraviolet space telescope capable of imaging terrestrial-sized planets orbiting in the habitable zones (HZ) of nearby stars, where abundant liquid water (and therefore life as we know it) could exist on the planet's surface. Understanding the HZ in terms of (1) its location around each star, and (2) the factors that could impact planetary climates within the HZ, is a longstanding area of research that will influence nearly every aspect of HWO's formulation (e.g. outer and inner working angles, starlight suppression capability, number of necessary targets, integration times, target follow-up, and mission duration).

Assessments of exoplanet habitability are typically based on first-order probes of whether a planet is capable of sustaining an atmosphere and liquid water on its surface, given the planet's mass, radius, equilibrium temperature, and the radiation environment of the host star \citep[e.g.][]{kas93,Abe2011,Leconte2013,kop13,kop14,Ramirez2014,shi14,Godolt2016,Kane2016,Wolf2017,Ramirez2018,Hill2023,Lobo2023,Spinelli2023}. Other planetary properties that affect the surface and atmosphere include the presence and strength of a magnetic field \citep[e.g.][]{Driscoll2013,Driscoll2015}; the structures and dynamical interplay of the surface and interior \citep[e.g.][]{Olson2006,Foley2016,Lenardic2016a,Lenardic2016b,Foley2018,Colose2021} and the bulk inventory of volatile and refractory elements and compounds \citep[for a review of the subject, see e.g.][]{Molliere2022}. All of these set the conditions for whether life can develop and survive over the lifetime of the planet. For the purposes of this paper, we draw from a definition referred to as ``climate habitability'', in the footsteps of works such as \citet{Jansen2019} and \citet{He2022}. This definition relies on a combination of surface temperature and precipitation and is described in detail in \S \ref{sec:habitability_models}, and is more narrowly defined than the requirement that surface liquid water be sustainable.

Our choice of climate habitability aims to build on this substantial body of work, by asking, ``For planets within the HZ that are similar to the Earth, how strongly could habitability be impacted by spin-orbit configuration?''. Specifically, we model climate habitability as a function of rotation rate, obliquity, eccentricity, and longitude of periastron. Changes to the orbital and rotation parameters factor greatly into the Earth's dramatic history of long term climate variability over millions of years \citep[e.g.][]{Milankovitch1941,ber93,spi10,Deitrick2018,Vervoort2022}. On short ($\sim$annual) timescales, seasonal variations are determined largely by the combination of rotation rate, obliquity, and orbital eccentricity \citep{Armstrong2014,Way2016,Way2023}. However, for more slowly rotating planets (i.e., day lengths longer than $\sim 1$/3 of a year), these factors combine to create a complex insolation pattern with significant inter-annual variations \citep[see e.g.][]{Dobrovolskis2013,Kane2016,Kane2017,Vervoort2022,Hill2023}.

We employ the ROCKE-3D code, which was also used by \citet{Jansen2019} and \citet{He2022}; more about our model setup is described in \S \ref{sec:atmospheric_models:GCM}. \citet{Jansen2019} start with an Earth-like initial condition for their climate models, and vary both the rotation period and insolation. Their principal finding is that the temperature- based fractional habitability has a well-defined peak in rotation period, with a rotation period of $\sim 16$ days having the highest $f_h$ for Earth-like insolation. They attribute this behavior to the efficiency of heat transport to the poles: the extent of Hadley circulation is determined by the Coriolis force, with circulation widening pole-ward as rotation slows from an Earth-like period. At rotation periods slower than $\sim 16$ days, habitability drops precipitously. The authors attribute this to the formation of sub-solar clouds; as the Sun lingers for longer periods of time in a given region on the globe, the persistent radiation draws more water into the atmosphere through evaporation. The models of \citet{He2022} span nearly the same range in rotation period as \citet{Jansen2019}, but add obliquity as a second dimension. \citet{He2022} also add a precipitation component to their working definition for habitability, which we adopt in a modified form. The full description and comparison of habitability definitions is available in \S \ref{sec:habitability_models:averaging}--\ref{sec:habitability_models:metric}.

Adding orbital eccentricity to the picture, which has both a magnitude and orientation, doubles the number of parameters to explore and therefore squares the size of a grid search of the parameter space. It is no longer feasible to simply span the space with climate models as we have been able to previously, and so we need a more efficient strategy to span the dimensions with a feasible number of model runs. This leads us into a statistical problem: given we can only sample this space sparsely, can we estimate how well we understand some outcome of our models --- such as a global and time mean surface temperature? We implement a combination of statistical methods: one to efficiently sample the parameter space (``Latin Hypercube'' sampling; see \S \ref{sec:atmospheric_models:grid}) and Bayesian statistical analysis in the form of Gaussian process regression (sometimes referred to as ``kriging''; see \S \ref{sec:habitability_models:emulating}).

We discuss the setup and configuration of the individual climate models, and how they are arranged in the rotation and orbital parameter space in \S \ref{sec:atmospheric_models}; definitions of habitability metrics and averaging in \S \ref{sec:habitability_models:averaging}--\ref{sec:habitability_models:metric}. We describe this statistical modeling and terminology in more detail in \S \ref{sec:habitability_models:emulating}. We compare the results of our coupled climate and statistical model with those of \citet{Jansen2019} and \citet{He2022} in \S \ref{sec:pre-results}, building up from rotation period alone, including obliquity, and finally including orbital eccentricity in \S \ref{sec:results}. As part of this, we design a test to evaluate how accurately the emulator predicts our habitability metric in \S \ref{sec:results:test}, by comparing the predictions of the emulator with the directly modeled habitabilities in the eccentric test model set. We interpret and discuss our results in \S \ref{sec:discussion}.

\section{Atmospheric Models}\label{sec:atmospheric_models}
\subsection{Configuration of the General Circulation Models (GCMs)}\label{sec:atmospheric_models:GCM}
For our climate simulations, we perform 93 runs with varied orbital parameters (obliquity, eccentricity, rotation rate, and longitude of periapse) with sampling as described in \S \ref{sec:atmospheric_models:grid} and values shown in Table \ref{table:ROCKE-3D_model_parameters}. This comprises two 46-model ensembles, which are used as ``training'' and ``test'' sets for purposes of the statistical analysis (as described in \S \ref{sec:habitability_models:emulating} and applied in \S \ref{sec:results:emulation}). Additionally, a model with Earth-like rotation and orbit parameters was run and added to the training set. The orbital and rotation configurations are listed in Table \ref{table:ROCKE-3D_model_parameters}, along with land and time means in surface temperature, precipitation, and our habitability metric values. All climate simulations presented employ ROCKE-3D \citep{Way2017}, a GCM developed at the NASA Goddard Institute for Space Studies. All models use the SOCRATES radiation scheme \citep{Edwards1996a,Edwards1996b,Thomson2019}. For this work, ROCKE-3D was configured with an atmosphere at $4^\circ \times 5^\circ$ latitude-longitude resolution, 40 vertical layers with a top at 0.1 mb, and a dynamic ocean with a uniform depth of 1360 m. This permits faster equilibration time than with a deeper ocean. Each model is run until the running mean across 30 model years (orbits) of the net bolometric radiative flux (\texttt{net\_rad\_planet} in the model outputs) is approximately zero. In practice, the typical variability in the net flux from year to year is of order a few $\sim 0.1$ W m$^{-2}$.\footnote{Assessing ``true'' climate equilibrium for Earth-derived climate models is a complex task, often without a clear-cut, unambiguous criterion; see e.g.~\citet{Rugenstein2020,DaiAiguo2020}.} The atmosphere uses Earth composition and pressure with 300 ppm CO$_2$ and 1 ppm CH$_4$. We do not simulate ozone chemistry in these experiments but prescribe a stratospheric ozone abundance that is hemispherically symmetric and seasonally invariant, allowing for the development of a stratospheric temperature inversion. We use an Earth continental configuration and a land surface with no vegetation. Instead, we use a bare soil mixture with a nominal surface albedo of 0.2, though this changes when the ground is wet or covered with snow. We choose to omit the modeling of ice sheets for our simulations given the extreme orbital configurations and significantly non-Earth-like insolation patterns. Despite this, the topographic heights are all still consistent with modern Earth. 

\subsection{A Parameter Study using Latin Hypercube Sampling}\label{sec:atmospheric_models:grid}
The grid of climate models presented here span a range in rotation period, obliquity, eccentricity, and longitude of periapse:
\begin{itemize}
    \item 8 powers of 2 in rotation period (denoted $P_\mathrm{rot}$), from 1--256 days;
    \item from 0--$90^\circ$ in obliquity (denoted $\psi$), spaced at intervals of $2^\circ$;
    \item from 0--0.225 in orbital eccentricity (denoted $e$), spaced at intervals of 0.005; and
    \item from 0--$360^\circ$ in longitude of periapse (denoted $\phi_\mathrm{p}$), spaced at intervals of $8^\circ$.
\end{itemize}
The maximum orbital eccentricity is limited to 0.225 to keep the climate models within a range that avoids numerical instability with the amount of instellation at periapse. This does not significantly limit the fraction these models explore of the currently-known exoplanet eccentricity distribution for planets with radii near Earth's, as constrained from transit surveys \citep[see e.g.][]{Kane2012b,VanEylen2015,An2023}.

\begin{deluxetable*}{rc||ccc|cc|ccc||ccc|cc|ccc}\label{table:ROCKE-3D_model_parameters}
\tabletypesize{\scriptsize}
\tablewidth{0pt}
\tablecaption{The orbital and rotation parameters used to initialize each of the training and test sets for the ROCKE-3D climate simulations, as well as the principal metrics for this paper averaged over time and land area --- see \S \ref{sec:habitability_models:averaging} for the precise definitions. $P_\mathrm{rot}$ is the planet's rotation period, $\psi$ its obliquity, $e$ its orbital eccentricity, and $\phi_\mathrm{p}$ its longitude of periapse. The metrics shown are the land and time mean surface temperature $\langle\bar{T}_\mathrm{s}\rangle$, precipitation $\langle\bar{p}\rangle$, and the 3 habitability metrics defined by temperature, precipitation, and a combination of the two (as defined in \S \ref{sec:habitability_models:metric}).}
\tablehead{\colhead{} &
           \colhead{} &
           \multicolumn{8}{c}{\textbf{Training Set}} &
           \multicolumn{8}{c}{\textbf{Test Set}} \\
           \colhead{Case} &
           \colhead{$P_\mathrm{rot}$ (d)} &
           \colhead{$\psi$ $\left(^\circ\right)$} &
           \colhead{$e$} &
           \colhead{$\phi_\mathrm{p}$ $\left(^\circ\right)$} &
           \colhead{$\langle\bar{T}_\mathrm{s}\rangle$ ($^\circ$C)} &
           \colhead{$\langle\bar{p}\rangle$ (mm/d)} &
           \colhead{$f_\mathrm{T}$} &
           \colhead{$f_\mathrm{prec}$} &
           \colhead{$H$} &
           \colhead{$\psi$} &
           \colhead{$e$} &
           \colhead{$\phi_\mathrm{p}$} &
           \colhead{$\langle\bar{T}_\mathrm{s}\rangle$} &
           \colhead{$\langle\bar{p}\rangle$} &
           \colhead{$f_\mathrm{T}$} &
           \colhead{$f_\mathrm{prec}$} &
           \colhead{$H$}
           }
\startdata
\textbf{Earth} & 1 & 23 & 0.017 & 283 & 8.0 & 2.85 & 0.69 & 0.81 & 0.57 &  -- & -- & -- & -- & -- & -- & -- & --  \\
\textbf{1} & 1 & 80 & 0.180 & 216 & 13.5 & 2.78 & 0.71 & 0.93 & 0.66 & 26 & 0.030 & 168 & 9.9 & 3.06 & 0.73 & 0.87 & 0.65 \\
\textbf{2} & 1.09 & 50 & 0.215 & 200 & 14.8 & 2.91 & 0.82 & 0.91 & 0.75 & 14 & 0.150 & 88 & 5.7 & 2.77 & 0.65 & 0.79 & 0.54 \\
\textbf{3} & 1.19 & 82 & 0.220 & 32 & 8.3 & 2.55 & 0.62 & 0.90 & 0.56 & 8 & 0.035 & 120 & 6.9 & 2.71 & 0.67 & 0.72 & 0.48 \\
\textbf{4} & 1.3 & 78 & 0.030 & 136 & 8.3 & 2.62 & 0.64 & 0.90 & 0.58 & 46 & 0.165 & 16 & 14.7 & 2.87 & 0.76 & 0.89 & 0.69 \\
\textbf{5} & 1.41 & 36 & 0.060 & 280 & 13.5 & 3.59 & 0.82 & 0.97 & 0.80 & 34 & 0.125 & 200 & 13.0 & 3.40 & 0.82 & 0.96 & 0.79 \\
\textbf{6} & 1.54 & 56 & 0.140 & 288 & 11.2 & 3.21 & 0.76 & 0.91 & 0.70 & 4 & 0.050 & 128 & 10.6 & 3.13 & 0.78 & 0.76 & 0.60 \\
\textbf{7} & 1.68 & 58 & 0.200 & 296 & 10.0 & 3.20 & 0.75 & 0.91 & 0.69 & 80 & 0.015 & 24 & 3.9 & 2.56 & 0.59 & 0.88 & 0.53 \\
\textbf{8} & 1.83 & 70 & 0.035 & 240 & 3.1 & 2.76 & 0.61 & 0.90 & 0.55 & 78 & 0.000 & 72 & 1.9 & 2.62 & 0.58 & 0.89 & 0.52 \\
\textbf{9} & 2 & 26 & 0.080 & 304 & 11.9 & 3.64 & 0.81 & 0.89 & 0.74 & 82 & 0.075 & 96 & $-1.5$ & 2.51 & 0.52 & 0.88 & 0.45 \\
\textbf{10} & 2.24 & 90 & 0.160 & 160 & $-3.6$ & 2.32 & 0.46 & 0.84 & 0.39 & 36 & 0.220 & 0 & 12.7 & 3.46 & 0.79 & 0.96 & 0.76 \\
\textbf{11} & 2.52 & 20 & 0.000 & 16 & 11.0 & 3.88 & 0.80 & 0.90 & 0.74 & 42 & 0.140 & 352 & 10.5 & 3.52 & 0.77 & 0.97 & 0.74 \\
\textbf{12} & 2.83 & 88 & 0.075 & 176 & $-4.8$ & 2.52 & 0.47 & 0.86 & 0.40 & 44 & 0.210 & 40 & 6.8 & 3.28 & 0.67 & 0.96 & 0.65 \\
\textbf{13} & 3.17 & 28 & 0.155 & 56 & 8.8 & 3.89 & 0.74 & 0.95 & 0.73 & 74 & 0.085 & 288 & $-2.5$ & 2.94 & 0.55 & 0.89 & 0.49 \\
\textbf{14} & 3.56 & 32 & 0.095 & 112 & 8.6 & 4.00 & 0.74 & 0.97 & 0.74 & 50 & 0.080 & 192 & 4.3 & 3.66 & 0.68 & 0.98 & 0.67 \\
\textbf{15} & 4 & 84 & 0.005 & 312 & $-5.4$ & 2.85 & 0.49 & 0.91 & 0.44 & 90 & 0.045 & 360 & $-5.6$ & 2.84 & 0.49 & 0.90 & 0.44 \\
\textbf{16} & 4.59 & 6 & 0.225 & 224 & 10.0 & 4.03 & 0.81 & 0.82 & 0.69 & 22 & 0.055 & 152 & 8.1 & 4.25 & 0.77 & 0.92 & 0.75 \\
\textbf{17} & 5.28 & 2 & 0.070 & 344 & 6.7 & 4.07 & 0.74 & 0.68 & 0.54 & 86 & 0.145 & 104 & $-8.4$ & 2.67 & 0.45 & 0.93 & 0.41 \\
\textbf{18} & 6.06 & 62 & 0.205 & 0 & $-2.7$ & 3.21 & 0.58 & 0.96 & 0.56 & 70 & 0.005 & 272 & $-4.5$ & 3.29 & 0.54 & 0.96 & 0.52 \\
\textbf{19} & 6.96 & 60 & 0.165 & 192 & $-1.3$ & 3.35 & 0.58 & 0.97 & 0.57 & 60 & 0.090 & 112 & $-2.1$ & 3.56 & 0.57 & 0.98 & 0.56 \\
\textbf{20} & 8 & 18 & 0.040 & 248 & 8.6 & 4.15 & 0.79 & 0.73 & 0.63 & 10 & 0.025 & 176 & 6.5 & 4.12 & 0.74 & 0.63 & 0.55 \\
\textbf{21} & 9.19 & 8 & 0.130 & 24 & 7.9 & 3.79 & 0.78 & 0.64 & 0.58 & 58 & 0.160 & 312 & 0.6 & 3.83 & 0.66 & 0.99 & 0.65 \\
\textbf{22} & 10.6 & 16 & 0.085 & 152 & 7.0 & 3.70 & 0.77 & 0.68 & 0.62 & 2 & 0.100 & 328 & 7.1 & 3.77 & 0.77 & 0.60 & 0.54 \\
\textbf{23} & 12.1 & 14 & 0.055 & 256 & 6.5 & 3.50 & 0.79 & 0.62 & 0.58 & 52 & 0.060 & 160 & 1.2 & 3.65 & 0.64 & 1.00 & 0.64 \\
\textbf{24} & 13.9 & 76 & 0.170 & 336 & $-6.3$ & 3.06 & 0.44 & 0.97 & 0.42 & 30 & 0.185 & 80 & 3.0 & 3.03 & 0.66 & 0.86 & 0.61 \\
\textbf{25} & 16 & 34 & 0.010 & 128 & 2.3 & 3.22 & 0.69 & 0.91 & 0.66 & 68 & 0.065 & 64 & $-8.1$ & 3.07 & 0.41 & 0.98 & 0.40 \\
\textbf{26} & 18.4 & 12 & 0.015 & 8 & 1.3 & 2.88 & 0.70 & 0.63 & 0.55 & 32 & 0.205 & 248 & 1.9 & 2.97 & 0.72 & 0.85 & 0.63 \\
\textbf{27} & 21.1 & 68 & 0.105 & 184 & $-9.8$ & 2.95 & 0.32 & 0.98 & 0.31 & 28 & 0.095 & 232 & $-0.2$ & 3.02 & 0.66 & 0.81 & 0.57 \\
\textbf{28} & 24.3 & 24 & 0.125 & 320 & $-2.7$ & 2.82 & 0.57 & 0.76 & 0.49 & 66 & 0.130 & 136 & $-11.8$ & 2.83 & 0.29 & 0.97 & 0.28 \\
\textbf{29} & 27.9 & 44 & 0.120 & 232 & $-7.3$ & 3.12 & 0.35 & 0.97 & 0.34 & 38 & 0.040 & 48 & $-7.0$ & 3.09 & 0.39 & 0.93 & 0.38 \\
\textbf{30} & 32 & 0 & 0.090 & 144 & $-6.9$ & 2.61 & 0.29 & 0.64 & 0.27 & 18 & 0.170 & 208 & $-7.0$ & 2.67 & 0.30 & 0.71 & 0.27 \\
\textbf{31} & 36.8 & 40 & 0.145 & 88 & $-12.3$ & 2.66 & 0.21 & 0.88 & 0.18 & 40 & 0.190 & 336 & $-10.7$ & 2.81 & 0.23 & 0.92 & 0.21 \\
\textbf{32} & 42.2 & 66 & 0.175 & 96 & $-16.9$ & 2.44 & 0.17 & 0.92 & 0.15 & 76 & 0.010 & 56 & $-14.2$ & 2.92 & 0.21 & 0.97 & 0.21 \\
\textbf{33} & 48.5 & 72 & 0.050 & 328 & $-14.6$ & 2.92 & 0.22 & 0.97 & 0.21 & 48 & 0.135 & 280 & $-12.9$ & 2.95 & 0.21 & 0.97 & 0.20 \\
\textbf{34} & 55.7 & 64 & 0.110 & 352 & $-15.5$ & 2.79 & 0.22 & 0.97 & 0.21 & 54 & 0.225 & 264 & $-14.0$ & 2.80 & 0.20 & 0.98 & 0.20 \\
\textbf{35} & 64 & 48 & 0.045 & 48 & $-16.0$ & 2.72 & 0.21 & 0.93 & 0.20 & 0 & 0.175 & 304 & $-14.0$ & 2.46 & 0.19 & 0.65 & 0.16 \\
\textbf{36} & 73.5 & 10 & 0.135 & 208 & $-15.2$ & 2.50 & 0.21 & 0.66 & 0.17 & 16 & 0.180 & 8 & $-15.7$ & 2.42 & 0.20 & 0.69 & 0.16 \\
\textbf{37} & 84.4 & 22 & 0.115 & 264 & $-16.1$ & 2.61 & 0.22 & 0.72 & 0.18 & 84 & 0.105 & 296 & $-15.7$ & 2.82 & 0.26 & 0.93 & 0.24 \\
\textbf{38} & 97 & 52 & 0.190 & 72 & $-19.4$ & 2.21 & 0.22 & 0.82 & 0.17 & 88 & 0.215 & 184 & $-17.7$ & 2.29 & 0.22 & 0.88 & 0.19 \\
\textbf{39} & 111 & 54 & 0.100 & 64 & $-18.2$ & 2.47 & 0.24 & 0.85 & 0.20 & 24 & 0.020 & 144 & $-17.4$ & 2.60 & 0.24 & 0.73 & 0.20 \\
\textbf{40} & 128 & 30 & 0.065 & 168 & $-18.2$ & 2.52 & 0.25 & 0.75 & 0.20 & 12 & 0.195 & 224 & $-18.2$ & 2.36 & 0.24 & 0.67 & 0.18 \\
\textbf{41} & 147 & 86 & 0.185 & 272 & $-16.6$ & 2.65 & 0.27 & 0.86 & 0.24 & 56 & 0.110 & 216 & $-17.4$ & 2.61 & 0.27 & 0.83 & 0.22 \\
\textbf{42} & 169 & 46 & 0.195 & 120 & $-19.9$ & 2.16 & 0.24 & 0.62 & 0.19 & 72 & 0.155 & 240 & $-16.9$ & 2.59 & 0.28 & 0.77 & 0.22 \\
\textbf{43} & 194 & 74 & 0.150 & 104 & $-18.8$ & 2.31 & 0.26 & 0.66 & 0.20 & 20 & 0.120 & 320 & $-18.5$ & 2.52 & 0.27 & 0.65 & 0.21 \\
\textbf{44} & 223 & 42 & 0.025 & 360 & $-19.0$ & 2.55 & 0.27 & 0.60 & 0.22 & 64 & 0.070 & 344 & $-18.0$ & 2.62 & 0.28 & 0.68 & 0.23 \\
\textbf{45} & 256 & 38 & 0.020 & 80 & $-19.5$ & 2.43 & 0.28 & 0.48 & 0.22 & 62 & 0.115 & 32 & $-19.2$ & 2.44 & 0.28 & 0.57 & 0.22 \\
\textbf{46} & 294 & 4 & 0.210 & 40 & $-22.1$ & 2.05 & 0.27 & 0.26 & 0.18 & 6 & 0.200 & 256 & $-22.0$ & 2.10 & 0.28 & 0.33 & 0.18 \\
\enddata
\end{deluxetable*}

The models for each set are spaced equally across each dimension's range except for rotation period which is divided logarithmically, in an approximately even grid, with finer divisions at the fast rotations. The division uses 8 steps between 1 and 2 day rotation period, 6 steps between 2 and 4 day rotation period, and 5 steps per factor of two for the remainder of the range. The result is a roughly logarithmic span over the full range. The maximum orbital eccentricity yields an orbit where the maximum insolation is 60\% and 150\% of its value at the semimajor axis at apoapse and periapse, respectively. Additionally, as described in the introduction, the longitude of periapse allows us to tune how the obliquity seasons are oriented in phase with the eccentricity seasons. For example, $\phi_\mathrm{p} \rightarrow 90^\circ$ means that the northern hemisphere reaches its summer solstice right as the planet hits periapse; $\phi_\mathrm{p} \rightarrow 270^\circ$ is the corresponding case for the southern hemisphere (and, therefore, the northern summer begins at apoapse).

A full lattice grid of models spanning these parameter ranges would take $N^4$ separate runs for $N$ models along each dimension. Since that would require us to run millions of models, we opt for a compromise where we run a much smaller set of models that is computationally feasible, but that ``spans'' the space as efficiently as possible with the available number of models. We employ \emph{Latin Hypercube} sampling (LHS) to span the space: on that lattice grid of $N$ equal spacings along each dimension, each of $N$ models sits on a vertex such that no other models share its dimension positions. The LHS technique has been widely used for efficiently exploring parameter spaces in general complex computer simulations. Our use of LHS was directly motivated by their use in exploring parameter spaces in Earth system models \citep[e.g.][]{Gregoire2010,Lee2011,Sexton2019,Kiang2021}. In order to ensure the specific sample set spans the space as evenly as possible, we selected the most evenly distributed sample set among a pool of 40 randomized candidate LHS sample sets generated by the \texttt{pyDOE} code \citep{pyDOE}. The quality of the distribution of samples within one LHS set is assessed by computing the pairwise euclidean distance between all possible sample pairs. We choose the candidate set whose ``worst case'' pairing --- the pair with the smallest separation in parameter space --- is the least ``bad'' (i.e. largest).

As we will discuss in \S \ref{sec:habitability_models:emulating}, we generate 2 independent samples of the parameter space using LHS; one to construct our model of the parameter space (the training set), and the other as a way to evaluate how well our model predicts outcomes in the spaces of the parameter space between the vertices it models directly (the test set). We calculate various averages of temperature and precipitation from the models for our habitability metrics as described below.

\section{Calculating Habitability}\label{sec:habitability_models}
\subsection{Defining Averages}\label{sec:habitability_models:averaging}
In presenting our results we consider multiple ways to average the outputs from the models. We define
\begin{itemize}
    \item a ``global'' average as the average of a quantity (such as surface temperature) over all grid cells, weighted by the surface area in each cell;
    \item a ``land'' average as similar to a global average, but where an additional weight is applied proportional to the fraction of land in each cell;
    \item a ``time'' average, which is the average over all model time outputs. Our most frequent model outputs are ``monthly'', which for eccentric orbits generalizes to 12 outputs roughly equally spaced in true anomaly. Since these will in general represent different time intervals, we weight the time average accordingly, yielding 360 time steps in total over 30 orbits (years) once each reaches the previously defined equilibrium condition;
    \item a ``total'' average as any spatial average (e.g. global or land-only), coupled with a time average; and
    \item a ``fractional'' average, which depends on the fraction of all time that each grid cell satisfies some criterion based on an output variable, such as having a monthly average surface temperature between 0 -- $100^\circ$C. This is done by assigning each grid cell a Boolean value of 0 or 1 per time step based on whether it satisfies the criterion, then taking a total average.
\end{itemize}

\subsection{Defining Habitability Metrics}\label{sec:habitability_models:metric}
The construction and analysis of our habitability metric are founded on two previous works. The first is \citet{Jansen2019}, who define their habitability metric based on a fractional area with a liquid water temperature ($0^\circ\leq T \leq 100^\circ$C):
\begin{gather}
\begin{aligned}\label{eq:Jansen_metric}
    f_h\!\left(P_\mathrm{rot}\right) &\equiv \frac{1}{A_\mathrm{tot}} \sum\limits_{i=1}^{n_\lambda} \sum\limits_{j=1}^{n_\phi} H_J\!\left(\lambda_i, \phi_j, P_\mathrm{rot}\right) A_{ij}, \\
    H_J\!\left(\lambda_i, \phi_j, P_\mathrm{rot}\right) &\equiv \begin{cases}
        1 & \mathrm{if}\quad 0^\circ \leq T\!\left(\lambda_i, \phi_j, P_\mathrm{rot}\right) \leq 100^\circ\,\mathrm{C} \\
        0 & \mathrm{else}
    \end{cases}
\end{aligned}
\end{gather}
for grid cells indexed in latitude $\lambda$ by $i$, longitude $\phi$ by $j$, with cell area $A_{ij}$, and for a model with rotation period $P_\mathrm{rot}$. In their work the habitability metric $H_J$ is evaluated on a temperature field that is averaged over at least one full orbital period. We note that, in both their models and the ones described here, since these models receive similar instellation values as Earth, the lower temperature threshold is much more important than the upper threshold, as detailed in our results (\S \ref{sec:results}).

We adopt the dual temperature-precipitation condition from \citet{He2022}, which combines the 0--$100^\circ$ C condition with a minimum yearly precipitation of 300 mm. This minimum value was originally chosen in \citet{He2022} to use Earth's major non-polar deserts as a reference for regions that receive insufficient rainfall \citep[using reference precipitation data from][]{Willmott2018}, despite generally having sufficiently warm temperatures . \citet{Spiegel2008} originally proposed the idea of the fractional habitability function --- based on the average fraction of global area capable of liquid surface water --- that could be integrated over arbitrary precision in time over one full orbital period. Since ROCKE-3D is a fully-3D code that requires significantly more computation time --- and produces significantly more output data --- than the 1D energy balance models (EBMs) used in \citet{Spiegel2008}, we are limited to sub-dividing our orbits into months. Any level of sub-annual sampling is important for eccentric orbits since the orbital rate, or the change in orbital position angle (known as true anomaly) can vary significantly across the orbit. The metrics we consider are fractional averages as defined above, and can be written in the form
\begin{gather}
\begin{aligned}\label{eq:fractional_land_averages}
    f_\mathrm{T}\!\left(P_\mathrm{rot}, \psi, e, \phi_\mathrm{p}\right) &\equiv \frac{1}{A_\mathrm{terr}} \sum\limits_{i=1}^{n_\lambda} \sum\limits_{j=1}^{n_\phi} \frac{1}{n_\mathrm{orb} P_\mathrm{orb}} \sum\limits_{\mathrm{month}=1}^{12 n_\mathrm{orb}} I_T\!\left(\lambda_i, \phi_j, t; P_\mathrm{rot}, \psi, e, \phi_\mathrm{p}\right) f_\mathrm{terr}A_{ij}\tau_k\!\left(e\right), \\
    f_\mathrm{prec}\!\left(P_\mathrm{rot}, \psi, e, \phi_\mathrm{p}\right) &\equiv \frac{1}{A_\mathrm{terr}} \sum\limits_{i=1}^{n_\lambda} \sum\limits_{j=1}^{n_\phi} \frac{1}{n_\mathrm{orb} P_\mathrm{orb}} \sum\limits_{\mathrm{year}=1}^{n_\mathrm{orb}} I_\mathrm{prec}\!\left(\lambda_i, \phi_j, t; P_\mathrm{rot}, \psi, e, \phi_\mathrm{p}\right) f_\mathrm{terr}A_{ij},
\end{aligned}
\end{gather}
where $I_\mathrm{X}$ functions as an indicator variable for a condition dependent on a variable X at any given grid cell indexed by $\left(\lambda_i, \phi_j\right)$ and time by $t$, sampled either by month or year/orbit. $n_\mathrm{orb}=30$ is the number of orbits from each run over which we average. $A_{ij} \propto \cos \lambda$ is the area of grid cell at $\left(\lambda_i, \phi_j\right)$, and $f_\mathrm{terr}$ is the fraction of that area that is the type of surface condition we would like to average over (i.e.~over only land, only ocean, or a global average, in which case $f_\mathrm{terr} \rightarrow 1$ and $A_\mathrm{terr} \rightarrow A_\mathrm{tot}=4\pi R^2_\oplus$). $\tau_k$ represents the variable month length across the orbit, since equal divisions in orbital anomaly are not equal intervals in time for an eccentric orbit. Our temperature indicator $I_\mathrm{T}$ is functionally identical to $H_J$ from Equation \ref{eq:Jansen_metric}, except that it is now also indexed in month subdivisions. $I_\mathrm{prec}$ returns 1 if the cumulative precipitation rate across an entire orbit is $\geq 300$ mm/year. We will refer to the climate habitability metric, or just habitability metric for the remainder of this work, as $H \equiv f_\mathrm{clim}$, where $I_\mathrm{clim} = I_\mathrm{T} \wedge I_\mathrm{prec}$. A model that satisfies the total yearly precipitation quota is automatically assigned a 1 for each month in that year.

\subsection{Emulating Habitability: A Multi-Dimensional Interpolation}\label{sec:habitability_models:emulating}
Our \emph{emulator} is an application of the ideas described in \citet{OHagan2006} and \citet{books/lib/RasmussenW06}. In the language of \citet{OHagan2006}, the grid of climate models we develop represent a \emph{simulator} of physical systems with observable outputs such as a global average temperature. Each 3-dimensional climate model represents our best estimator of that observable at that particular set of rotation and orbital parameters, but is computationally expensive to run. To estimate a model observable at arbitrary points in the parameter space, we could choose to interpolate between the directly modeled points, creating a function that represents our best estimate of the observable as a function of the parameters. The process of \emph{emulation} ultimately provides an interpolation, but also builds a model of the uncertainty in our knowledge of the remaining parameter space --- a statistical model of the variable. This approach resembles the sparse sampling and emulation process used in Earth system modeling to explore model outputs over ranges of parameter space \citep{Lee2011}. Note that this general sampling and emulation approach is widely used in Earth science for uncertainty estimation \citep{Uusitalo2015} and model calibration \citep{Fletcher2022}, although different studies can use a variety of specific sampling or emulation techniques.

A common way to construct such a statistical model is to use \emph{Gaussian processes}, which \citet{OHagan2006} and \citet{books/lib/RasmussenW06} also describe in detail. This sort of emulation is finding a growing application in multiple fronts of exoplanet modeling \citep[see e.g.][]{Haqq-Misra2024a}. Gaussian processes assume that the probability distribution for an ``observed'' quantity --- say, mean surface temperature on a planet with Earth-like topography and ocean cover at a given rotation and orbital configuration --- is approximated reasonably by a Gaussian (normal) distribution. The uncertainty in a global temperature is averaged over multiple model orbits, and so will have uncertainties from inter-annual variation as well as the inherent limitations of the computer simulation in approximating the real physical system. This produces a posterior distribution for the value of a measurable at any valid point in parameter space, whose prior is determined by some set of parameters (set by a ``kernel'', detailed later) that describe the overall model. In other words, this is a Bayesian framework where prior knowledge is encoded in how we choose our kernel function. This whole procedure is known as Gaussian process ``regression'' --- an interpolation in $N$ dimensions with an estimate of the statistical uncertainty. A convenient outcome of this model approach is that the uncertainty of the interpolation function itself can then be described as a multivariate normal distribution.

Building the emulator is therefore a matter of choosing a function, composed of kernel functions, that describes how the mean and uncertainty of the distributions of the average temperature, precipitation, or habitability at each point in space should vary. An emulator will always choose the mean of its distribution for the actually-modeled points as whatever those models return, and there will be no statistical uncertainty in the emulator at those points. There may be uncertainty in model observables due to the aforementioned factors such as inter-annual variability and computational uncertainty, but here we refer to the \emph{additional} uncertainty in the prediction relative to what we observe in the climate models, since we assume that every time one runs the climate model at that rotation and orbital configuration, over the same time frame, one will always return the same average temperature. To build our emulators, we use the \texttt{scikit-learn} package in Python \citep{scikit-learn}. The two kernel functions that we explore in this work are
\begin{itemize}
    \item the \emph{radial basis function}, or RBF, which only depends on the distance between two points,
\begin{equation}\label{eq:RBF_kernel}
k\!\left(\mathbf{x}_a, \mathbf{x}_b\right) = \exp\left[-\frac{\sum_{i=1}^{4} \left(x_{ai}-x_{bi}\right)^2}{2\ell^2}\right]
\end{equation}
where $\mathbf{x}_a$ and $\mathbf{x}_b$ are two arbitrary points whose coordinates in our 4-dimensional space are indexed by $i$; and
\item a white kernel function:
\begin{equation}\label{eq:white_kernel}
k\!\left(\mathbf{x}_a, \mathbf{x}_b\right) = \sigma \delta_{ab}
\end{equation}
for some constant noise level $\sigma$ and the Delta function $\delta_{ab}$ of the positions of the two points. In other words, the white kernel contains no correlated noise by construction; the only correlation will be modeled through the RBF.
\end{itemize}
A single RBF in our composite kernel means there is one length scale per dimension that represents the typical length scale on which the observable varies. The white kernel allows the emulator to estimate the average additional amount of statistical uncertainty in the data when fitting with the radial basis function kernel.

Finally, since this construction assumes a Cartesian-like coordinate system for our parameter space, we recast our eccentricity variables $e$ and $\phi_\mathrm{p}$ in sines and cosines of the longitude, i.e.~$\left(e, \phi_\mathrm{p}\right) \rightarrow \left(e \cos\phi_\mathrm{p}, e \sin\phi_\mathrm{p}\right)$. The reason is that these two variables on their own behave more like a polar coordinate system, with the eccentricity representing a magnitude and the longitude of periapse representing an angle.

One way to check the predictive quality of an emulator is to run additional climate models at a new set of points, thereby directly simulating the outputs at points previously only estimated by the emulator, and compare those outputs with the distributions predicted by the emulator. If the emulator is well ``tuned'', the predicted values and their uncertainties should encompass the directly-simulated habitability values. Conversely, if the emulator fails to reliably predict the outcomes of the new set of model outputs, it may indicate, for example, that a quantity (such as mean global temperature) varies with one or more of the parameters (such as rotation period or eccentricity) on finer scales than could be predicted from the previous set of models. Borrowing language from machine learning, we call the outputs from the initial set of climate models the ``training'' data, in that they are mock observations of climates used to train our emulator, and the new climate models are the ``test'' data.

In preparation of emulating the new grid of climate models, we start by re-examining temperature, precipitation, and habitability for non-eccentric climate models, and compare our findings with existing results. This is partly to benchmark our models with previous efforts, and also to have a baseline to distinguish the effects of each new dimension as they are added to the model grid.

\section{Applying the Habitability Metrics to Existing Models}\label{sec:pre-results}
\subsection{Comparison with Previous Habitability Metrics}\label{sec:pre-results:rotation-only}
\begin{figure}[htb!]
\begin{center}
\includegraphics[width=17cm]{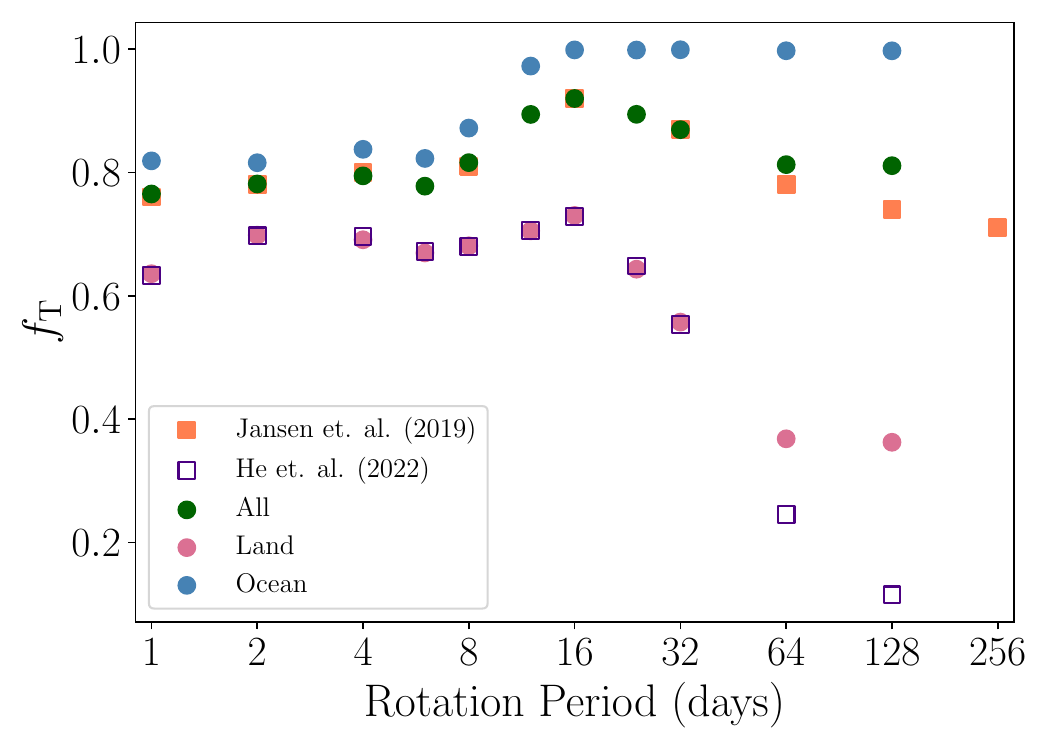}
\caption{A comparison of the global temperature-based habitability metric of \citet{Jansen2019}, in orange squares, as defined in Equation \ref{eq:Jansen_metric}, with the fractional land temperature averages ($f_\mathrm{T}$) in circles, as calculated from Equation \ref{eq:fractional_land_averages}. The fractional averages are colored based on whether they are averaged over the entire globe (green), only the oceans (blue), or only land (pink). Our global fractional averages agree very closely with the results of \citet{Jansen2019} to a maximum rotation period of 32 days, beyond which our values remain constant out to the model grid maximum of 128 days while the Jansen values continue to decrease. This is due to the differences in annual vs. monthly averaging; see \S \ref{sec:pre-results:rotation-only} for more discussion. The original calculation of the fractional averages over land in \citet{He2022} are shown as unfilled purple squares, whose averaging approach was similar to that in \citet{Jansen2019}.}
\label{fig:zero-obliquity_comparison}
\end{center}
\end{figure}

We start by recalculating the temperature metric $f_\mathrm{T}$ for simulations over a range of rotation periods at zero obliquity, to compare to the results from \citet{He2022} and \citet{Jansen2019}. While all three studies use  ROCKE-3D, with only minor differences in the model setups, each study computes the average habitability metrics in different ways. This produces significant differences in the results, even when the temperature criteria remain the same (between 0$^\circ$ and 100$^\circ$C). In \citet{Jansen2019}, the temperature habitability was computed from the annually averaged near surface atmospheric temperature (using the 984 mb pressure level) over the entire global area. In \citet{He2022}, the focus was shifted to habitability over land surfaces, so the quantities were computed only over the land-covered grid cells in the simulation. Since \citet{He2022} use simulations over a range of obliquity, seasonal variations are important and will produce habitable regions in isolated space--time regions. For example, polar regions on Earth have habitable periods during the summer even though the annual average temperature could be less than 0$^\circ$C. To account for this effect, the habitability was computed over a monthly climatology averaged over the last 30 years of simulation time (i.e. a running 30-year average of January, of February, etc.), after the model had reached radiative equilibrium. While this method will correctly capture regional and seasonal zones of habitability, it will not correctly capture regional zones of habitability driven by the long diurnal cycles in the slowly rotating planets because the diurnal cycle is not in-phase with the orbital cycles. Since the two cycles are out of phase, over long averaging periods (e.g. the 30 years) the diurnal cycle will tend to average out to zero with the monthly climatology. For the slowest rotators (e.g., 128 and 256 rotation period) the diurnal cycle itself can create habitable temperatures for long periods (at time scales of $\sim$ months for the present-day Earth) that could be considered ``growing seasons'' even though they are driven by rotation instead of obliquity. Therefore, in the present study, we compute the average temperature habitability by applying the threshold procedure on the full time series of monthly average temperatures and then averaging the result over a number of orbits to reduce the impact of weather variability. This is clearly shown in Equation \ref{eq:fractional_land_averages}, where the indicator variable $I_X$ is computed before averaging over time (sum over index $k$).

In Figure \ref{fig:zero-obliquity_comparison}, we show a comparison between the temperature habitability metrics from \citet{Jansen2019} and \citet{He2022} and results from applying our new method as described above on the \citet{He2022} simulations. First, the \citet{Jansen2019} results are nearly identical to our results for the global average (``All'' in Figure \ref{fig:zero-obliquity_comparison}) for rotation periods less than 64 days. Here, the only differences we expect are small, due to small differences in the exact ROCKE-3D configuration or machine floating point round-off errors. The vast majority of the structure seen is due to temperatures dropping below $0^\circ$ C, as temperatures exceeding $100^\circ$ C are very rare. At 64 and 128-day rotation periods, the divergence is due to the difference in averaging methods. For any grid cell where the temperature through the diurnal cycle crosses 0$^\circ$C part of the time, its contribution to the total fractional habitability will be larger with our method compared to the \citet{Jansen2019} method if the annual mean temperature is less than 0$^\circ$C, or smaller if the average temperature is larger than 0$^\circ$C. In the global average, it is clear that the contribution from grid cells with annual temperature less than 0$^\circ$C have more effect since our global temperature metric is increased for the two slow rotation periods. Second, comparing to the \citet{He2022} results, we see the habitability is overall lower, because the habitability is computed from only the land grid cells. In general, the ocean surface is warmer overall due to the larger heat capacity and the ability to transport heat horizontally through ocean currents. The differences between the \cite{He2022} results and our new calculations follow the same trend as described above in the comparison to \citet{Jansen2019}.

For all the monthly-averaged cases using our new method, the 64-day and 128-day cases appear to have nearly identical values of $f_\mathrm{T}$. This result implies that while the contrast in absolute temperature between the dark and bright times of the year may widen, the fraction of the year spent above freezing is remaining nearly constant. The ocean averages in fact maximize at values of nearly 100\% starting at a rotation period of 16 days, and remain there through a rotation period of 128 days. This due to the higher heat capacity and thermal transports in the oceans as mentioned above. Meanwhile, the land averages decrease sharply starting at $P_\mathrm{rot}=16$ days, as the periods of extended darkness lengthen. Land surfaces are not nearly as capable of retaining thermal energy as the oceans are and therefore the temperatures drop much lower during periods with no insolation.

It is with this in mind that we focus primarily on the land-only fractional averages for our main analysis in the remainder of this work. The land averages capture the response of the most sensitive parts of the globe to changes in insolation brought on by changes in the rotation and orbital parameters, particularly with regard to land plants which comprise roughly 80\% of Earth's biomass \citep{Bar-On2018} and contribute significantly to the reflected light spectrum of Earth \citep[e.g.][]{Seager2005,Arnold2009,OMalley-James2018,Schwieterman2018}. Moving from variations in rotation period alone, we add obliquity as a second dimension, rounding out the rotation parameters and extending the work presented in \citet{He2022}.

\subsection{Emulating across the He et. al. (2022) Model Grid}\label{sec:pre-results:He-emulation}
\begin{figure}[htb!]
\begin{center}
\includegraphics[width=17cm]{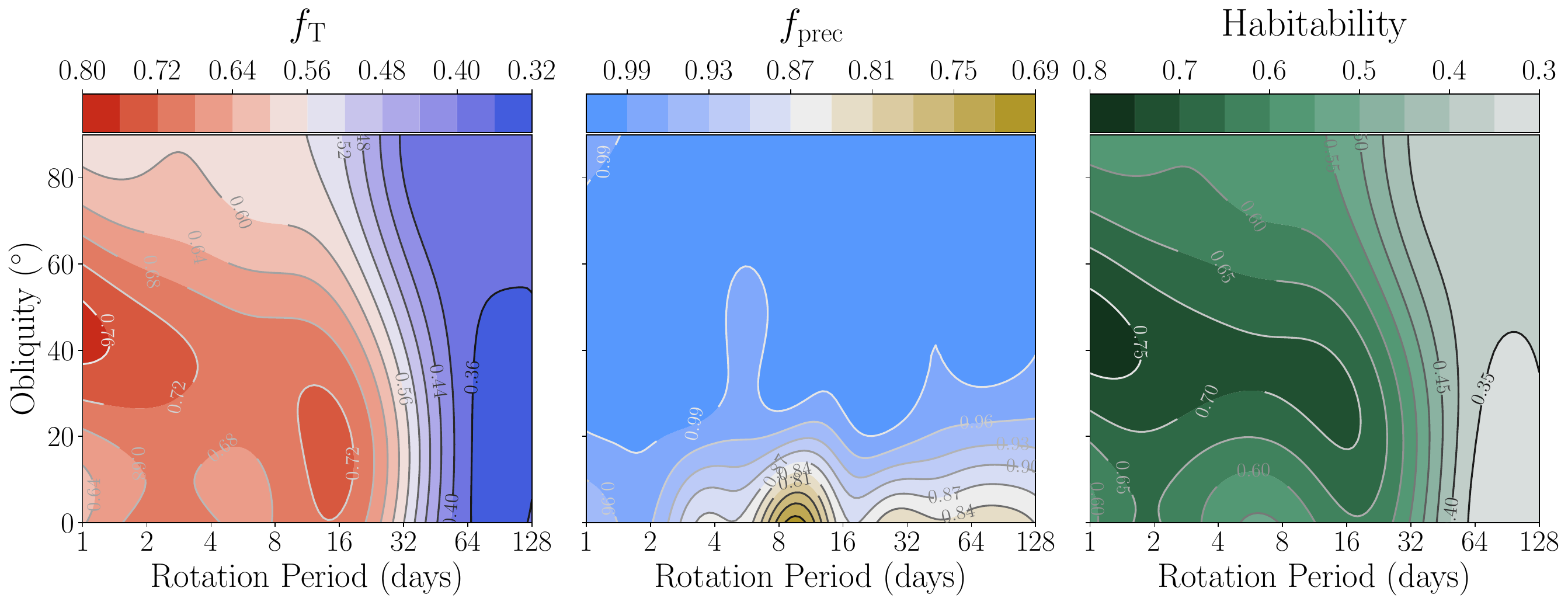}
\caption{Contours denoting the emulated values of the fractional land average temperature $f_\mathrm{T}$, precipitation $f_\mathrm{prec}$, and habitability metric $H$, as calculated from Equation \ref{eq:fractional_land_averages} and defined in \S \ref{sec:habitability_models:metric}, for the climate models originally published in \citet{He2022}. For non-eccentric orbits, rotation period is the strongest predictor of whether on average a planet's land will maintain a minimum surface temperature and therefore habitability. The precipitation metric drops only at obliquities near zero at isolated rotation periods, which are attributed to features of the Hadley circulation distributing precipitation more or less evenly in latitude.}
\label{fig:He-grid_emulation}
\end{center}
\end{figure}
Figure \ref{fig:He-grid_emulation} shows the results of applying our emulation technique to the original grid of models used in \citet{He2022} (compare with for example Figure 8 in their work). Table \ref{table:kernel_parameters} shows the associated fit values of the kernel parameters ($\ell$ and $\sigma$, as defined in Equations \ref{eq:RBF_kernel} and \ref{eq:white_kernel}, respectively). This time, we focus on fractional averages over land, versus over the entire globe. Familiar patterns emerge in the fractional temperature averages, with a sharp negative gradient in temperature with rotation periods beyond $\sim 32$ days. The peaks in temperature-based habitability ($f_\mathrm{T}$) are seen at rotation periods $\sim 1$ day and obliquities 30--$70^\circ$, as well as the previously noted peak at rotation periods close to 16 days and low obliquities. The precipitation-based habitability $f_\mathrm{prec}$, being considered over an entire orbit, is much more uniform across rotation period and obliquity. This metric will not distinguish between a planet that receives an even amount of precipitation over one orbit, versus one that receives the same amount of precipitation but concentrated in a short time window within each orbit. We find the overall range of $f_\mathrm{prec}$ is higher, and only dips below $\approx 0.9$ for isolated regions at or near zero obliquity. This can be attributed to the dependence of the extent of Hadley circulation with rotation rate, which controls the latitude and extent of deserts.

\section{Statistics across the Grid of Eccentric Climate Models}\label{sec:results}
\subsection{Global Temperature Averages across the Eccentric Grid}\label{sec:results:global-averages}
\begin{figure}[htb!]
\begin{center}
\includegraphics[width=17cm]{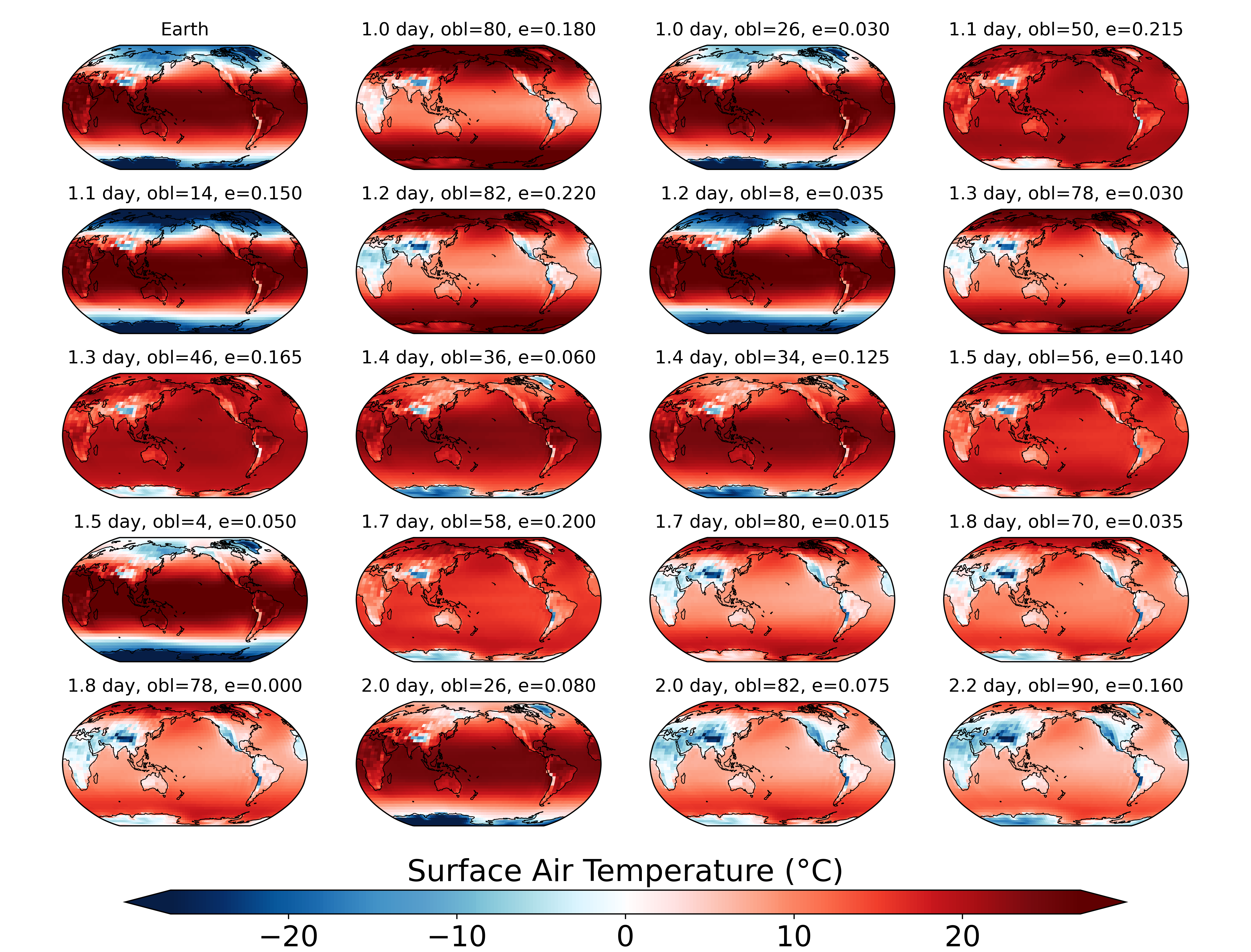}
\caption{Time averages of the surface temperature for the 1st quintile of the training and test models in rotation period, ordered fastest to slowest. While the rotation periods are organized in increasing order from top left to bottom right, the remaining spin and orbital parameters (obliquity, eccentricity, and longitude of periapse) vary according to the configuration from the Latin Hypercube Sampling algorithm, and are themselves not in any particular order.}
\label{fig:surface_temperature_averages_set1}
\end{center}
\end{figure}

\begin{figure}[htb!]
\begin{center}
\includegraphics[width=17cm]{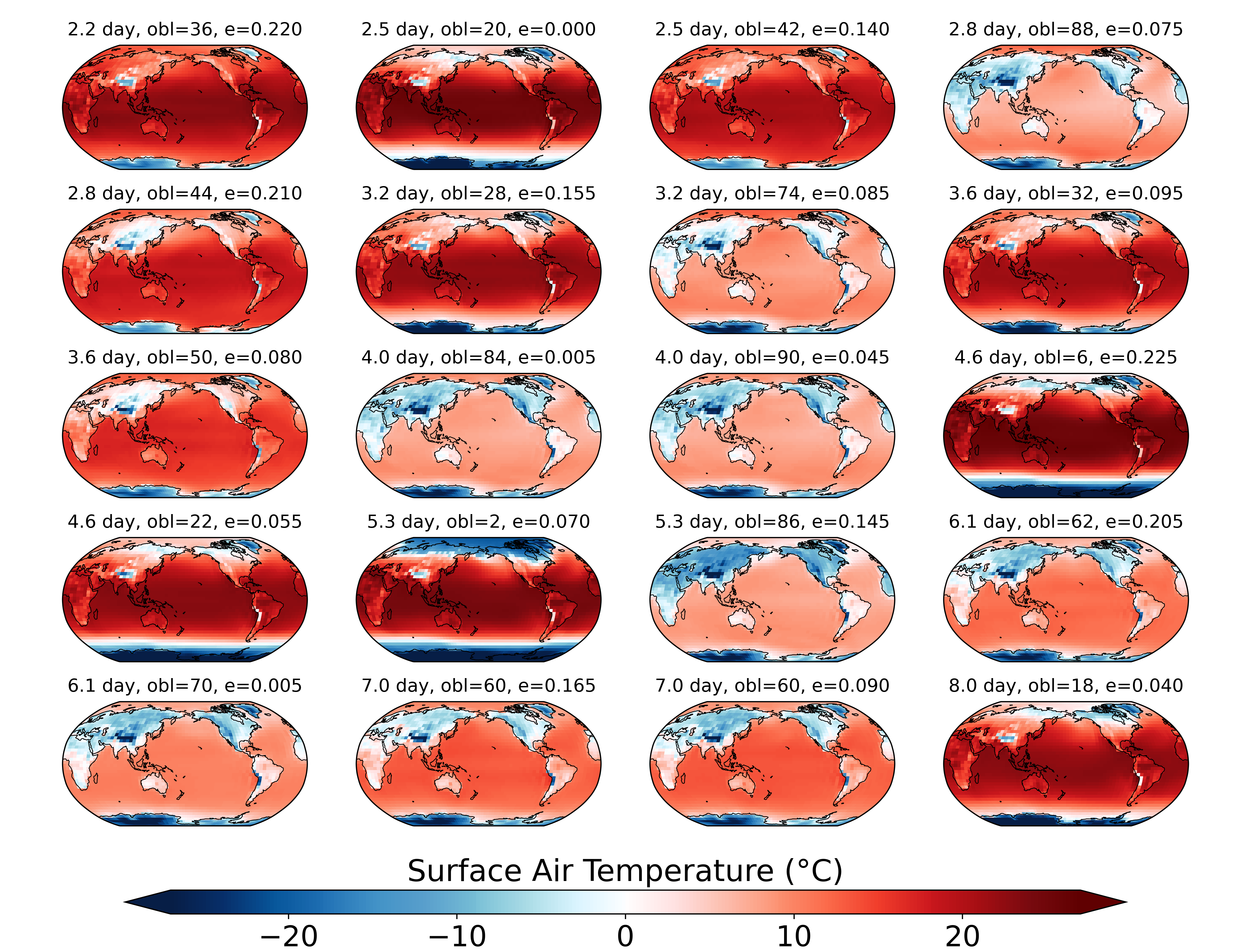}
\caption{Time averages of the surface temperature for the 2nd quintile of the training and test models in rotation period, ordered fastest to slowest. While the rotation periods are organized in increasing order from top left to bottom right, the remaining spin and orbital parameters (obliquity, eccentricity, and longitude of periapse) vary according to the configuration from the Latin Hypercube Sampling algorithm, and are themselves not in any particular order.}
\label{fig:surface_temperature_averages_set2}
\end{center}
\end{figure}

\begin{figure}[htb!]
\begin{center}
\includegraphics[width=17cm]{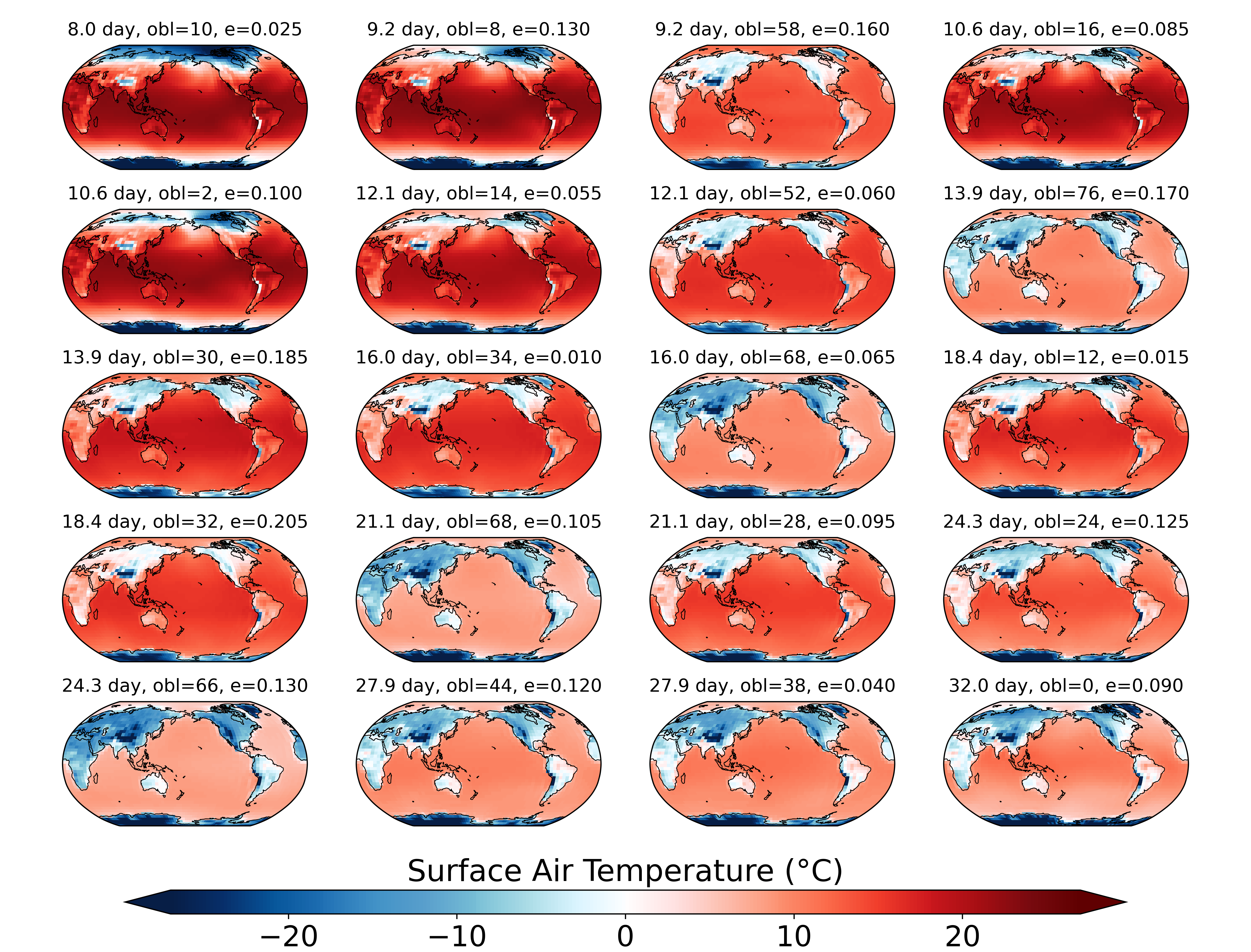}
\caption{Time averages of the surface temperature for the 3rd quintile of the training and test models in rotation period, ordered fastest to slowest. While the rotation periods are organized in increasing order from top left to bottom right, the remaining spin and orbital parameters (obliquity, eccentricity, and longitude of periapse) vary according to the configuration from the Latin Hypercube Sampling algorithm, and are themselves not in any particular order.}
\label{fig:surface_temperature_averages_set3}
\end{center}
\end{figure}

\begin{figure}[htb!]
\begin{center}
\includegraphics[width=17cm]{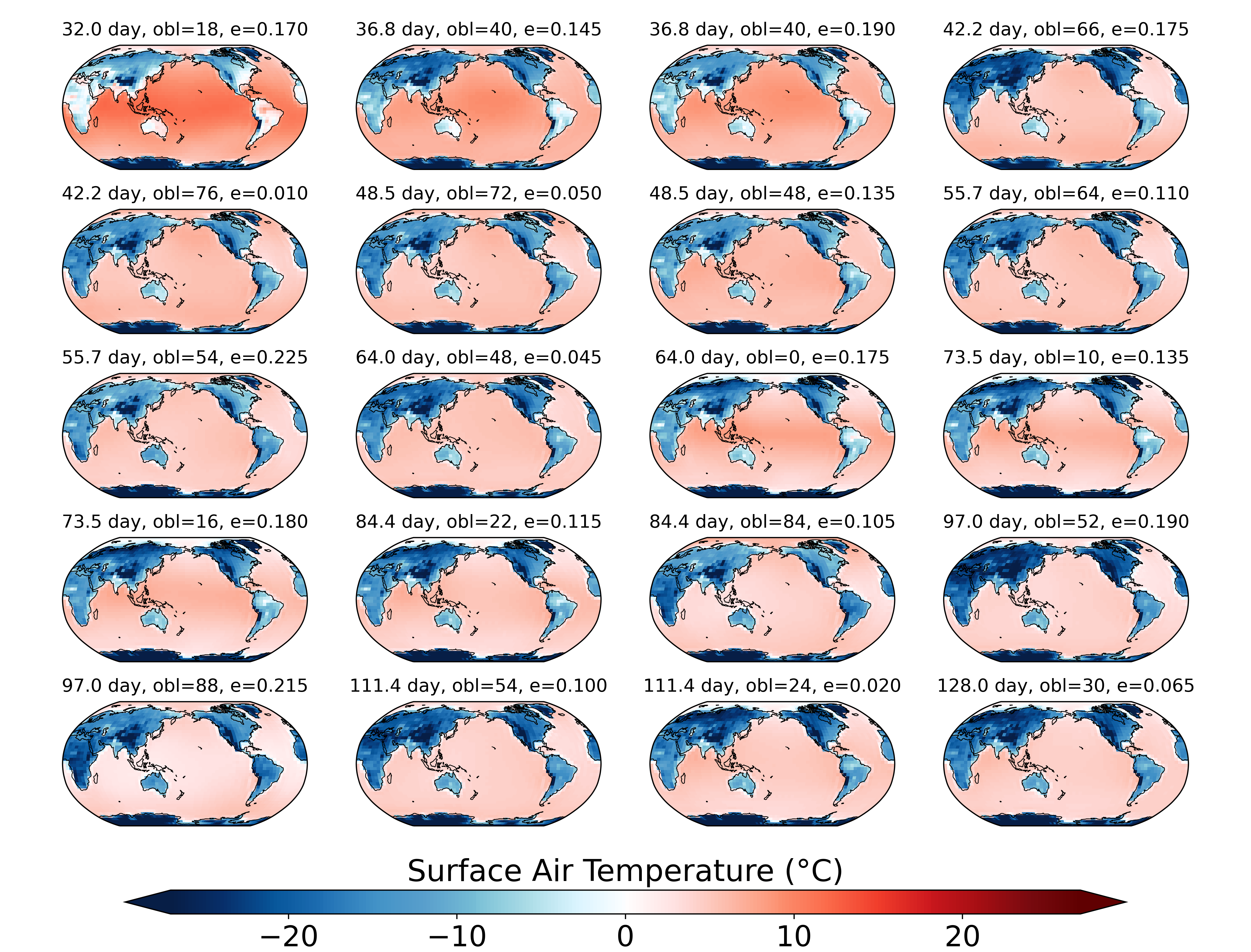}
\caption{Time averages of the surface temperature for the 4th quintile of the training and test models in rotation period, ordered fastest to slowest. While the rotation periods are organized in increasing order from top left to bottom right, the remaining spin and orbital parameters (obliquity, eccentricity, and longitude of periapse) vary according to the configuration from the Latin Hypercube Sampling algorithm, and are themselves not in any particular order.}
\label{fig:surface_temperature_averages_set4}
\end{center}
\end{figure}

\begin{figure}[htb!]
\begin{center}
\includegraphics[width=17cm]{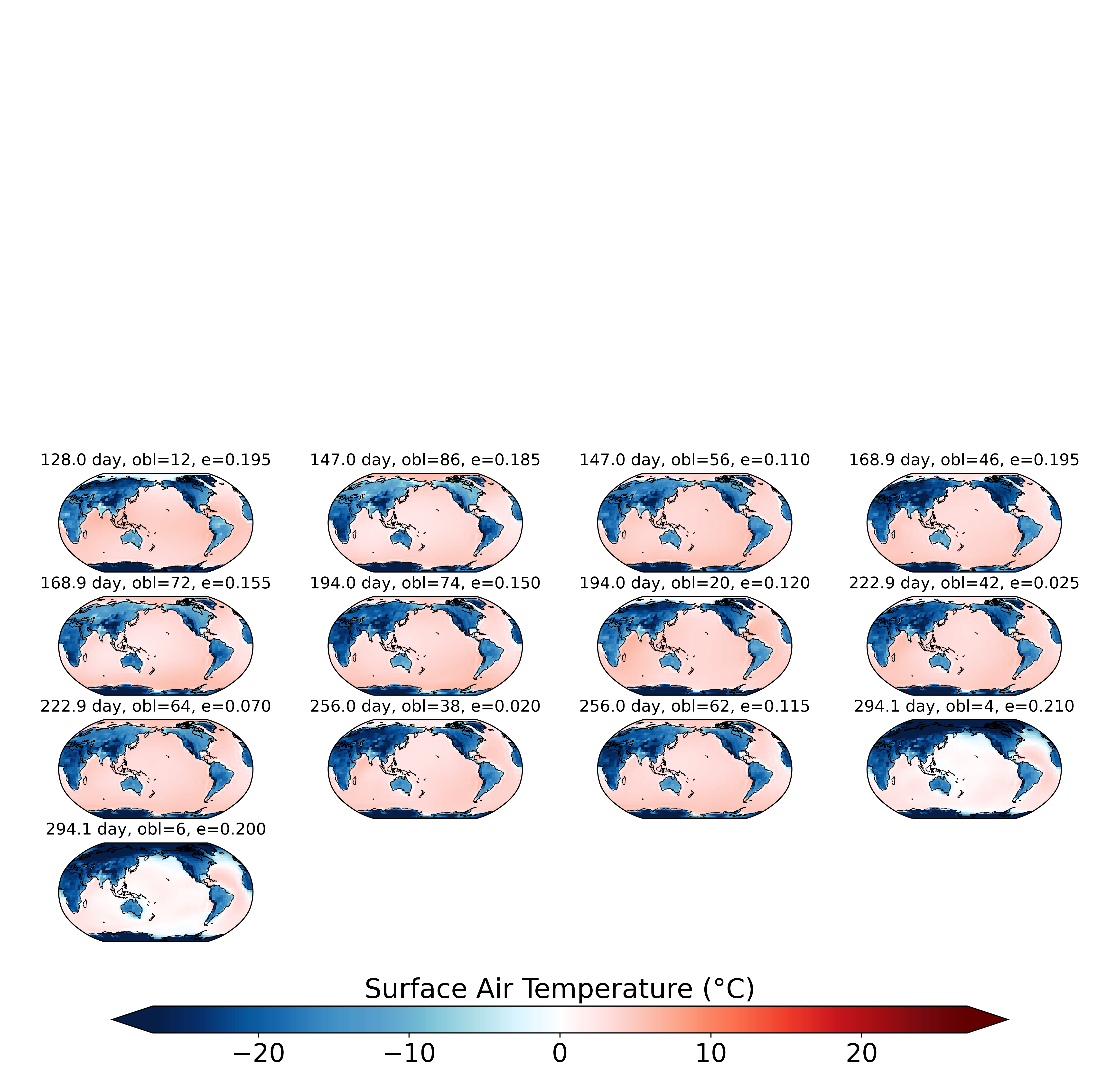}
\caption{Time averages of the surface temperature for the 5th quintile of the training and test models in rotation period, ordered fastest to slowest. While the rotation periods are organized in increasing order from top left to bottom right, the remaining spin and orbital parameters (obliquity, eccentricity, and longitude of periapse) vary according to the configuration from the Latin Hypercube Sampling algorithm, and are themselves not in any particular order.}
\label{fig:surface_temperature_averages_set5}
\end{center}
\end{figure}

Figures \ref{fig:surface_temperature_averages_set1}--\ref{fig:surface_temperature_averages_set5} show the climatological surface temperature pattern for the simulations in the training and test ensembles, ordered by rotation period. At the fastest rotation periods, we can classify the patterns of global and time mean surface temperatures into 3 main types:
\begin{description}
\item[Low obliquity yields cold poles] At low obliquities, there is a broad region in latitude centered at the equator with temperatures significantly above $0^\circ$. Regions near the poles show a marked contrast with mean temperatures significantly below zero.
\item[Moderate obliquity yields warm oceans] At moderate obliquities, we see the smallest temperature contrast across the entire globe, as stellar energy is distributed the more evenly in latitude. Regions near the poles are now significantly warmer, sometimes well above freezing, with the coldest regions less severely below zero. These coldest regions are limited to continental interiors near the poles (e.g.~the interiors of Greenland and Antarctica), and high elevations such as the Himalayan and Andean mountain ranges.
\item[High obliquity yields hot poles] At the highest obliquities, we see the hottest regions at the poles, followed by cooler but still warm oceans nearer the equator. The interiors of non-polar continents are now either near or slightly below freezing on average, with the coldest temperatures at the highest elevations. In these cases, the interiors of the Canadian and Siberian continental plains, as well as Antarctica, are among the warmest regions on the planet.
\end{description}
The last of these three types, the ``hot poles'' type, only persists for rotation periods of 1--2 days. These are the only models that are seen to reach the upper temperature threshold of $100^\circ$ C, with the model with the most frequent occurrence seeing such temperatures for 1.7\% of its grid cells across latitude, longitude, and time. The low obliquity cases continue to provide warm temperatures across continents and oceans near the equator, with the moderate obliquity cases evolving more of a cold-continent versus warm-ocean contrast. These two types exist until the rotation periods slow to $\sim 20$ days; at this point, low obliquities are no longer able to maintain an equator-to-pole temperature contrast, and the continent-ocean temperature contrast takes over for all obliquities. The ``limiting'' state of global and time mean surface temperatures at slow rotations is a global mean ocean temperature just above freezing, with all continents --- regardless of latitude --- averaging tens of degrees Celsius below freezing.

While we have focused here on discussing rotation period and obliquity, the grids of models also include variations in orbital eccentricity up to $e=0.225$. However, in the multi-year averages these have a relatively insignificant effect on global temperatures, as we show further in the following sections.

\subsection{Temperature, Precipitation, and Habitability Metrics}\label{sec:results:metrics}
\begin{figure}[htb!]
\begin{center}
\begin{tabular}{c}
\includegraphics[width=17cm]{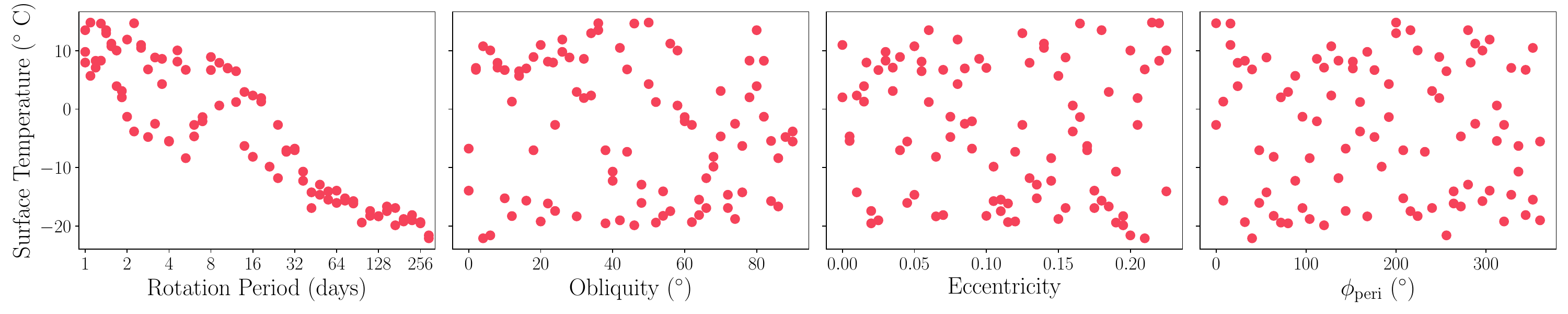} \\
\includegraphics[width=17cm]{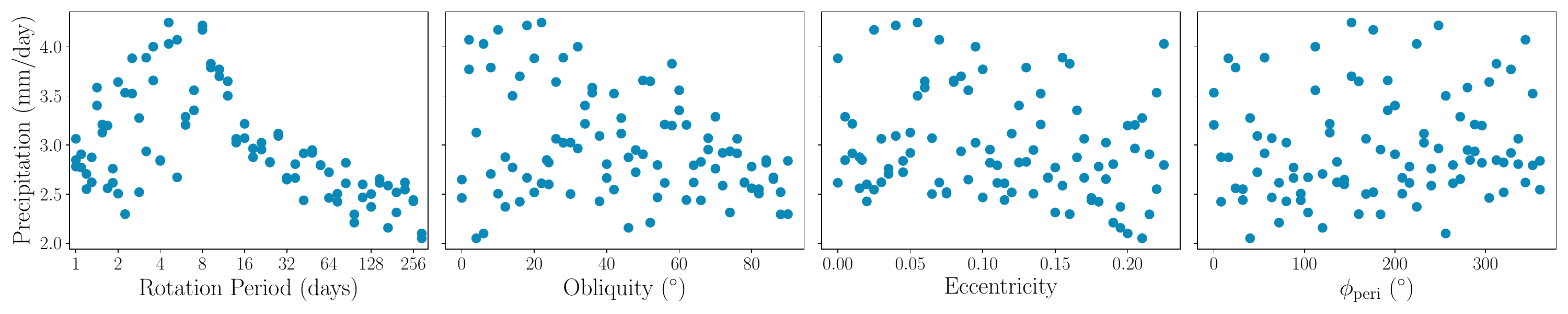} \\
\end{tabular}
\caption{The total land averages of surface temperature and precipitation, each as a function of the 4 varied parameters of the model runs. The average temperature generally is negatively correlated with rotation period, while the other parameters show very little structure. Average precipitation exhibits a peak at rotation periods of several days and at low obliquities, though it should be noted that, on this scale, the average land precipitation rates are well above our specified threshold for habitability ($\approx 0.82$ mm/day = 300 mm/year).}
\label{fig:total_land_average_training_scatterplots}
\end{center}
\end{figure}

We begin by plotting the total land average of the surface temperature as a function of each of the two rotation and two orbital parameters (Figure \ref{fig:total_land_average_training_scatterplots}). As with previous results with non-eccentric model grids, the average surface temperature depends strongly on the rotation period. The total land averaged temperature first dips below $> 0^\circ$C for a case with a rotation period just above 2 days; at intermediate rotation periods ($2 < P_\mathrm{rot} < 20$ days) a substantial fraction of the runs dip below freezing; and $\gtrsim 20$ days all runs average $< 0^\circ$C. Additionally, we see there is a hint of a bifurcation in the region faster than 32 days, with an ``upper track'' of points remaining above $0^\circ$C out to $\sim 20$ days, and the ``lower track'' dipping below freezing beyond a rotation period of 2 days. In contrast, the obliquity and eccentric parameters show very little correlation with these temperature averages. Obliquity is responsible for the aforementioned bifurcation in average land temperatures for rotation periods faster than $\approx 32$ days. However, we also see that there is a lower limit as well: at rotation periods faster than about 2 days, the apparent correlation with obliquity is much weaker.

\begin{figure}[htb!]
\begin{center}
\includegraphics[width=17cm]{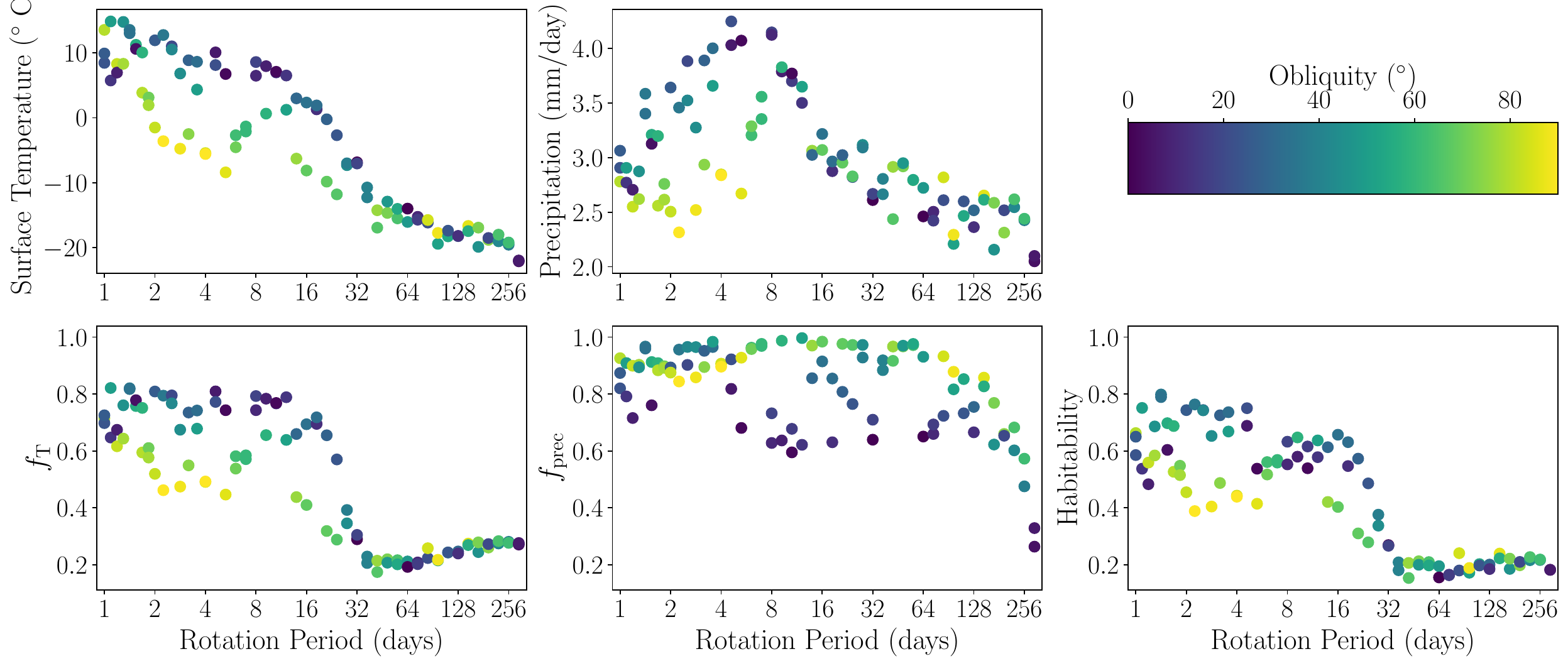}
\caption{Fractional averages of the surface temperature (left) and precipitation (center) for each of the model runs. Fractional averages are defined as the averages over all model times, weighted by area, of land that satisfies the prescribed habitability conditions in temperature ($0<T<100^\circ$C) and precipitation (precipitation $>300$ mm/year). By this construction, the habitability metric (right) is a fractional average with the combined conditions.}
\label{fig:fractional_average_scatterplots}
\end{center}
\end{figure}

Moving to the precipitation rates, we see some different structure in both rotation period and obliquity. In rotation period we see a broad peak in the average precipitation at rotation periods of 4--8 days, with a subsequent drop beyond $\gtrsim 10$ days. In obliquity, the very highest average precipitation rates ($\gtrsim 3.5$ mm/day) occur only for cases with $\psi < 30^\circ$. And, as with surface temperature, we see a dependence in obliquity for fast rotation periods (upper right panel of Figure \ref{fig:fractional_average_scatterplots}), with the range where the contrast is highest between 1--8 days. For reference, the Earth's real average precipitation is $\sim 2.7$ mm/day across all continents and seasons, and an average of 3.5 mm/day would be comparable to many humid regions in Earth's subtropics, including many cities along the Gulf of Mexico and Atlantic coasts of the United States such as Houston, Texas; Savannah, Georgia; and New York City. Regardless of the model, all precipitation averages exceed the $\approx 0.85$ mm/day minimum used for our cutoff for habitability, a rate which is much closer to a semi-arid locale such as Los Angeles or much of the interior of Australia.

To prepare our analysis of the habitability metric, we can instead look at the average fraction of land area that satisfies \emph{either} the minimum temperature or precipitation in a given month. In Figure \ref{fig:fractional_average_scatterplots} we show the fractional land averages of temperature ($f_{\mathrm{T}} = f_{\mathrm{T}>0^\circ\mathrm{ C}}$) and precipitation ($f_{\mathrm{prec}} = f_{\mathrm{prec}>300\,\mathrm{ mm/yr}}$), as well as the habitability metric values, which by definition is the combination of these two conditions. The structure of $f_\mathrm{T}$ is similar to that of the total averages in temperature, with a slight flattening of the previously noted upper track and a reversal of the slope seen at the longest ($\gtrsim 32$ day) rotation periods. We highlight in color tracks that are distinguished by obliquity: between rotation periods of 2--32 days, both rotation period and obliquity have strong influences on the temperature habitability.

The fractional average of precipitation also shows a potential bifurcation, with obliquities at or above roughly 30$^\circ$ staying at or above 0.8 until rotation periods $\gtrsim 100$ days. The lowest obliquities drop to values around 0.6, with again little dependence on rotation period until $\gtrsim 100$ days. The contrast seen in both the non-eccentric (middle panel of Figure \ref{fig:He-grid_emulation}) and eccentric precipitation habitability for low obliquity cases is modest compared with the drop-off in metric values for the slowest rotation periods. When comparing the average precipitation to the precipitation metric values, we find a striking contrast: for rotation periods $\lesssim 12$ days, low obliquity cases tend to have higher average precipitation across land area and time, but \emph{lower} metric values than their higher obliquity counterparts. This implies that, on low-obliquity fast rotators, precipitation is plentiful, but concentrated in certain regions of land; on higher-obliquity fast rotators, less precipitation is received on average, though enough to generally satisfy our minimum assumed threshold, and it is much more equitably spread out across land area.

Combining these in the habitability metric, we see the ``two regimes'' behavior is underscored: a quickly-rotating regime where both rotation period and obliquity determine habitability, and a drop-off to slower rotations where habitability is roughly constant out to our maximum simulated rotation periods. In the combined habitability metric, the structure of the fractional temperature metric dominates, and the contrasting slopes of the temperature and precipitation metrics at rotation periods beyond 32 days combine to produce a roughly constant habitability with rotation at the slowest values.

\subsection{Emulator Results: Interpolating between the Training Data}\label{sec:results:emulation}
\begin{deluxetable}{rc|ccccc}[htb]\label{table:kernel_parameters}
\tabletypesize{\footnotesize}
\tablewidth{0pt}
\tablecaption{Fit parameters for the kernel of the Gaussian process regressors (emulators) used to interpolate the habitability metrics. The length and noise scales $\ell$ and $\sigma$ are defined in Equations \ref{eq:RBF_kernel} and \ref{eq:white_kernel}, respectively.}
\tablehead{
    \colhead{} &
    \colhead{} &
    \multicolumn{4}{c}{RBF Length Scales ($\ell$)} &
    \colhead{White Noise Level ($\sigma$)} \\
    \colhead{Model Set} &
    \colhead{Metric} &
    \colhead{$\log_2\!\left(P_\mathrm{rot}/\mathrm{days}\right)$} &
    \colhead{$\psi$ $\left(^\circ\right)$} &
    \colhead{$e\cos\phi_\mathrm{p}$} &
    \colhead{$e\sin\phi_\mathrm{p}$} &
    \colhead{}
    }
    
\startdata
\multirow{3}{*}{\textbf{Non-eccentric}} & $f_\mathrm{T}$    & 1.34 & 31.6 & & & $1.67\times10^{-3}$ \\
                                        & $f_\mathrm{prec}$ & 0.56 & 21.7 & & & $3.12\times10^{-3}$ \\
                                        & $H$               & 1.29 & 31.7 & & & $5.94\times10^{-3}$ \\
\hline
\multirow{3}{*}{\textbf{Training}} & $f_\mathrm{T}$    & 1.26 & 50.9 & 1.41 & $\gtrsim 10^{3}$ & $\lesssim 10^{-9}$ \\
                                   & $f_\mathrm{prec}$ & 1.45 & 33.7 & 0.54 & $\gtrsim 10^{3}$ & $8.8\times10^{-3}$ \\
                                   & $H$               & 1.20 & 40.3 & 1.63 & $\gtrsim 10^{3}$ & $\lesssim 10^{-9}$ \\
\hline
\multirow{3}{*}{\textbf{Training+Test}} & $f_\mathrm{T}$ & 1.17 & 56.7 & 1.28 & $\gtrsim 10^{3}$ & $\lesssim 10^{-9}$ \\
                                   & $f_\mathrm{prec}$   & 1.28 & 30.4 & 0.84 & 1.92 & $\lesssim 10^{-9}$ \\
                                   & $H$                 & 1.18 & 43.4 & 1.54 & 3.81 & $\lesssim 10^{-9}$ \\
\enddata
\end{deluxetable}

We have made the case that, based on the analysis of the total land and fractional averages, we should see our interpolations show the greatest structure in rotation period, followed by obliquity, and finally by the eccentric parameters. To visualize all 4 dimensions on a 2-dimensional page, we construct a ``grid-of-grids'': the outermost axes span two of the dimensions, and each inner plot shows the remaining two dimensions with the outer two dimensions fixed to values according to where the inner plot lies in the grid. Here we choose to plot the eccentric dimensions on the outer axes defining the grid, and each plot within the grid shows the rotation dimensions: rotation period on the $x$ axis, obliquity on the $y$ axis. For the outer axes the eccentric dimensions are also recast as $x_\mathrm{outer} = e\cos\phi_\mathrm{p}$, $y_\mathrm{outer} = e\sin\phi_\mathrm{p}$. By doing this, we represent the eccentricity dimensions in a polar form: the absolute eccentricity of each inner plot corresponds to its distance from the ``center'' of the grid, and the longitude of periapse corresponds to the angular position around the center of the grid.

\begin{figure}[htb!]
\begin{center}
\begin{tabular}{c}
\includegraphics[height=10.5cm]{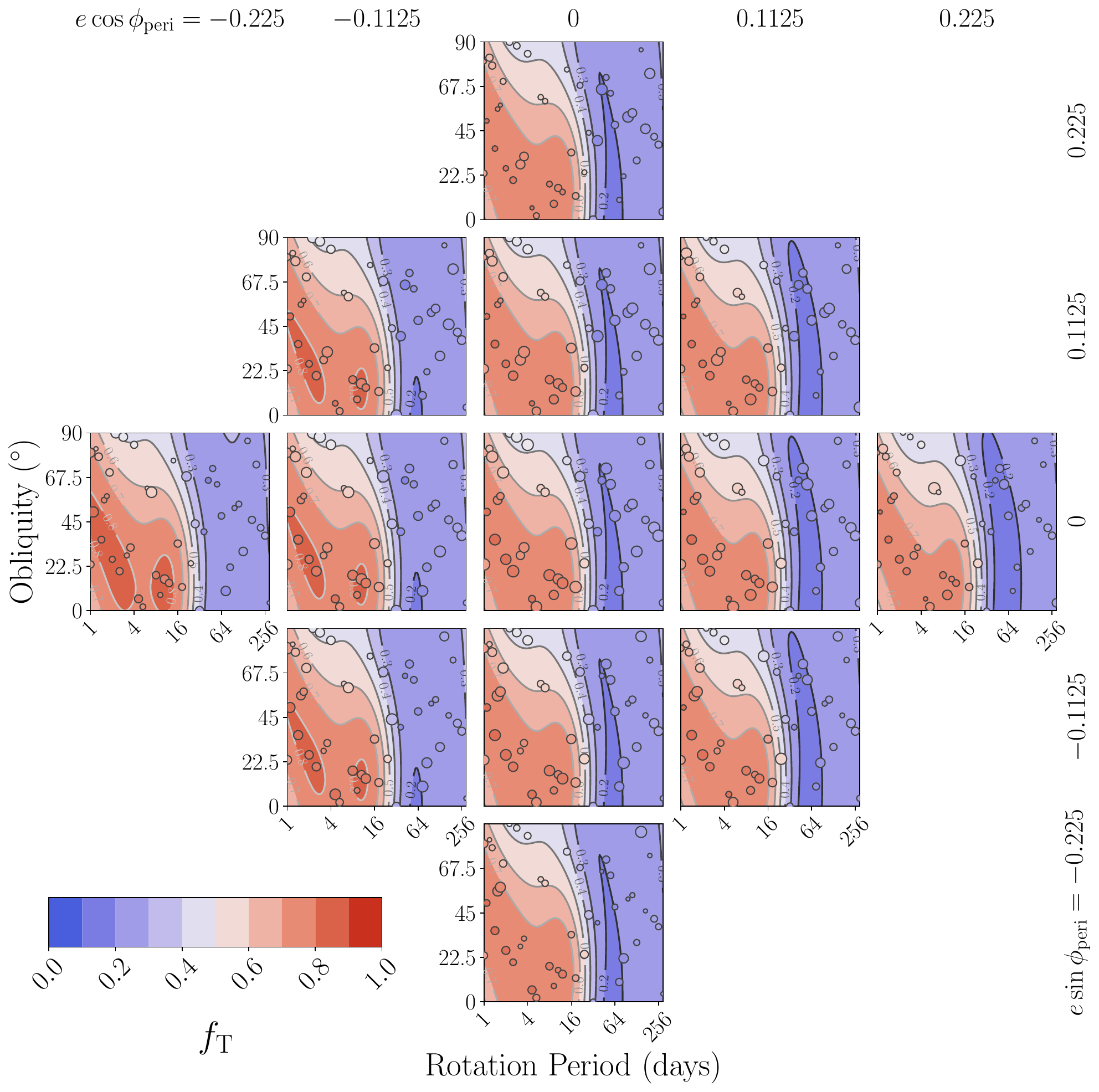} \\
\includegraphics[height=10.5cm]{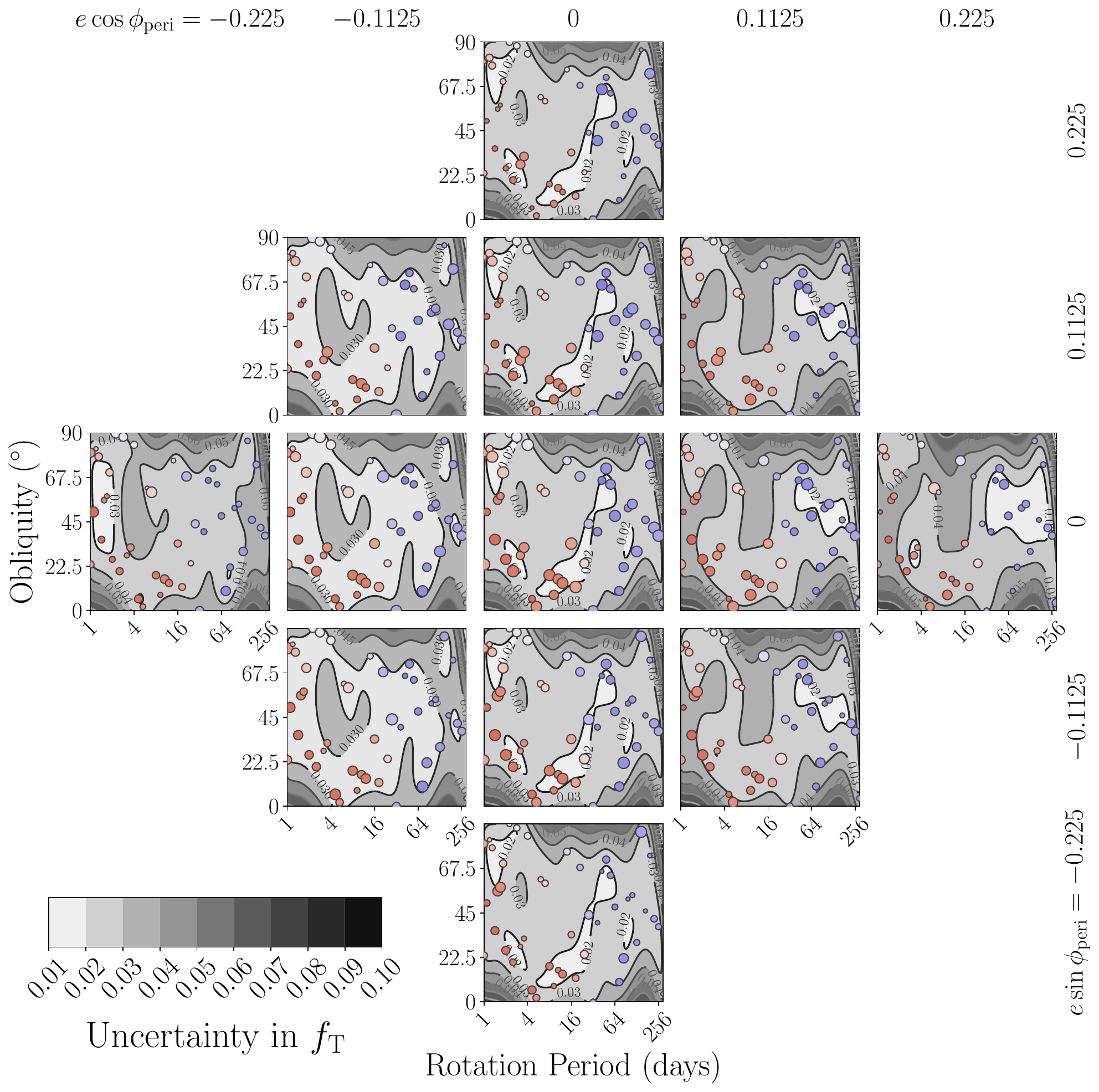} \\
\end{tabular}
\caption{The emulated fractional averages of surface temperature ($f_\mathrm{T}$) for the training set. The emulations are shown across the four model dimensions as a ``grid of grids'', where each sub-plot shows rotation period versus obliquity, and the sub-plots are arranged such that eccentricities increase radially from the center sub-plot. The outcomes of the actual runs are over-plotted at their locations as points, colored on the same scale. Each sub-plot contains all model points, but since each sub-plot is fixed at specific values of eccentricity and longitude of periapse, the points' apparent sizes are scaled to represent their ``distance'' in eccentricity from their projected position on that sub-plot. On the lower plot we show the corresponding uncertainties in the predictions as grey-scale filled contours, with higher uncertainties as darker shades.}
\label{fig:SurfaceTemperature_training_grid-of-grids}
\end{center}
\end{figure}

\begin{figure}[htb!]
\begin{center}
\begin{tabular}{c}
\includegraphics[height=10.5cm]{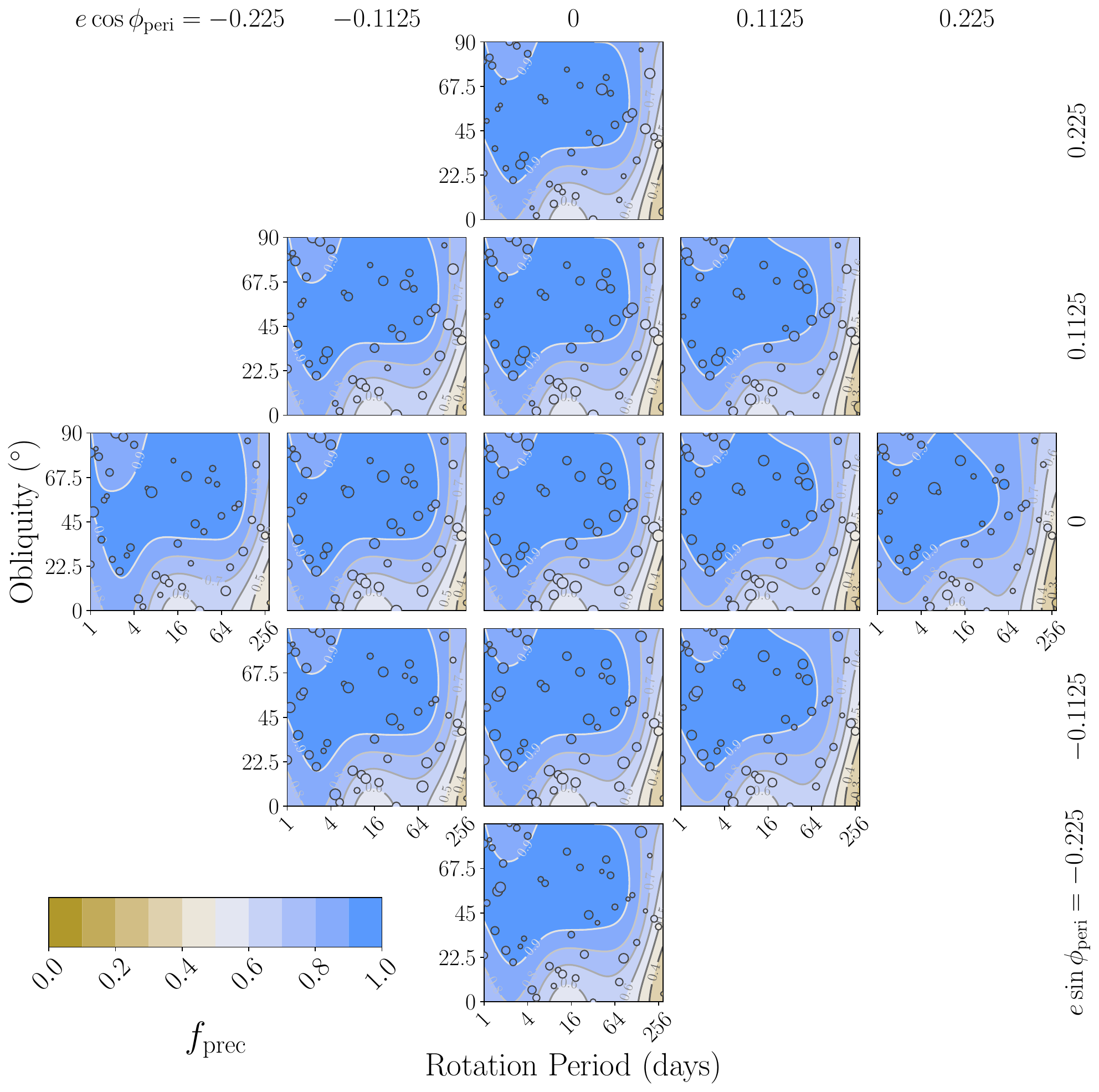} \\
\includegraphics[height=10.5cm]{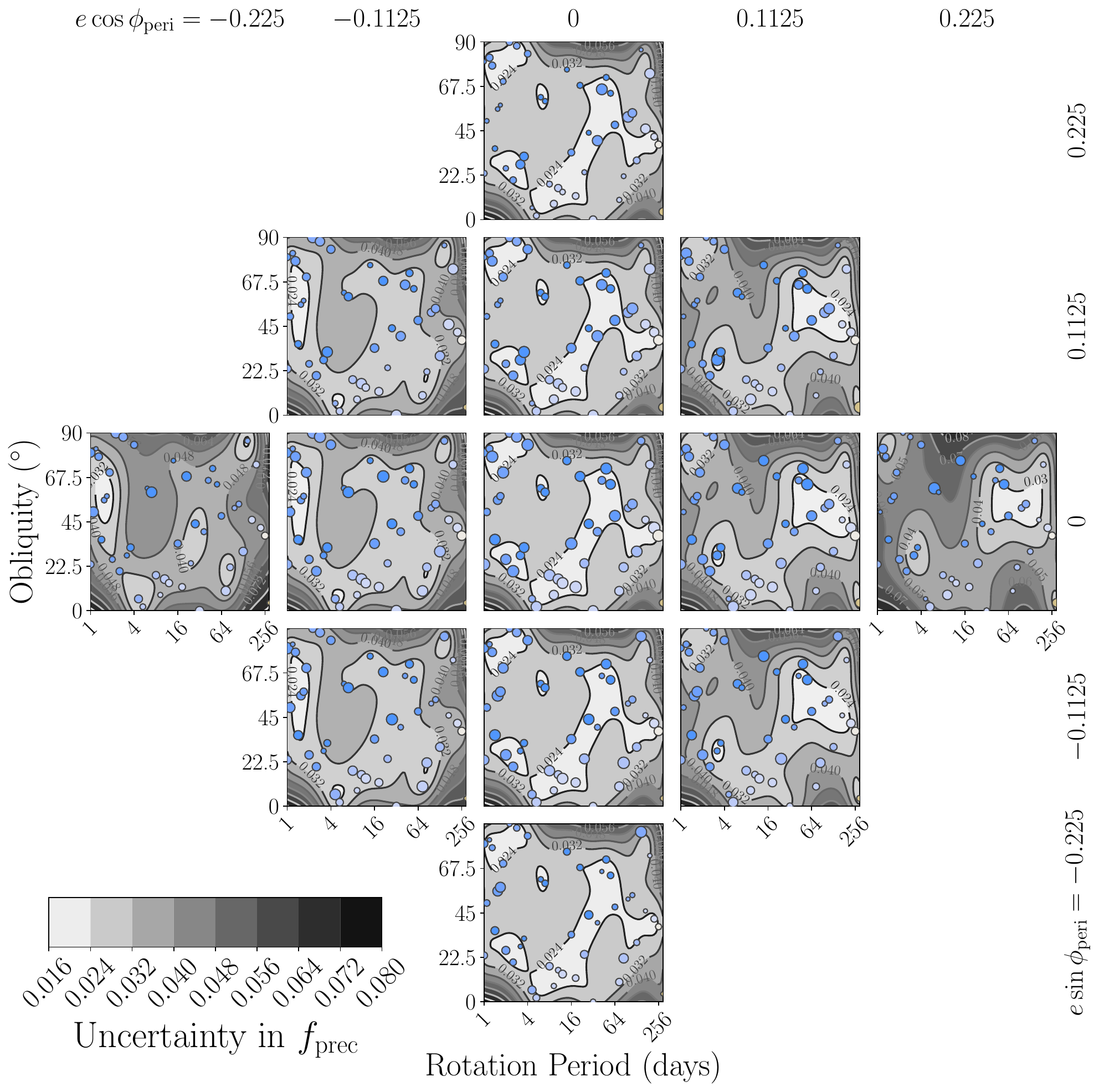} \\
\end{tabular}
\caption{The emulated fractional averages of precipitation ($f_\mathrm{prec}$) for the training set. The emulations are shown across the four model dimensions as a ``grid of grids'', where each sub-plot shows rotation period versus obliquity, and the sub-plots are arranged such that eccentricities increase radially from the center sub-plot. The outcomes of the actual runs are over-plotted at their locations as points, colored on the same scale. Each sub-plot contains all model points, but since each sub-plot is fixed at specific values of eccentricity and longitude of periapse, the points' apparent sizes are scaled to represent their ``distance'' in eccentricity from their projected position on that sub-plot. On the lower plot we show the corresponding uncertainties in the predictions as grey-scale filled contours, with higher uncertainties as darker shades.}
\label{fig:Precipitation_training_grid-of-grids}
\end{center}
\end{figure}

We start by expanding the view of our fractional average temperatures (Figure \ref{fig:SurfaceTemperature_training_grid-of-grids}) and precipitation (Figure \ref{fig:Precipitation_training_grid-of-grids}) into what their emulators predict across the parameter space. Our choice of a radial basis function to interpolate between our training model points means that the contours form regions of low uncertainty around clusters of points; each sub-plot shows a 2-dimensional ``slice'' into these 4-dimensional regions. In the surface temperature slices we see that the emulator draws a plateau of high temperate fractions spanning two edge regions of the parameter space: from rotation periods of a few days at near-zero obliquities, to the quickest ($\sim 1$-day) rotation periods at intermediate ($\approx 45^\circ$) obliquities. This plateau is strongest as one moves to the leftmost column, i.e.~where eccentricity is significant, and where the longitude of periapse is in the region centered at $\approx 180^\circ$. The previously noted correlations of temperature with rotation period and obliquity are reflected clearly in the interpolations, with a peak gradient between rotation periods $\sim 16$--32 days regardless of eccentricity. We also see a shallower gradient in obliquity for rotation periods between approximately 2--16 days. The uncertainties in the predicted temperatures are tied to the absolute distance, in our 4-dimensional parameter space, from our training points; regions of low predicted uncertainties radiate from clusters of points, mimicking the features seen in the predicted temperatures.

In precipitation, we see less contrast due to the annual nature of the metric. There is a drop-off in $f_\mathrm{prec}$ at all eccentricities that occurs roughly around rotation periods of 100 days, with a slight dependence on obliquity. The lowest metric values occur at the lowest obliquities and longest rotation periods, and the emulator finds a slight dependence in this corner of the parameter space with $e\cos\phi_\mathrm{p}$. However, this eccentricity dependence hinges on just one point in this area; it is more likely that the emulator chooses to ``return to the mean'' value as we move to the opposite side of $e\cos\phi_\mathrm{p}$ values, rather than there being robust evidence for a physical phenomenon.

\begin{figure}[htb!]
\begin{center}
\begin{tabular}{c}
\includegraphics[height=10.5cm]{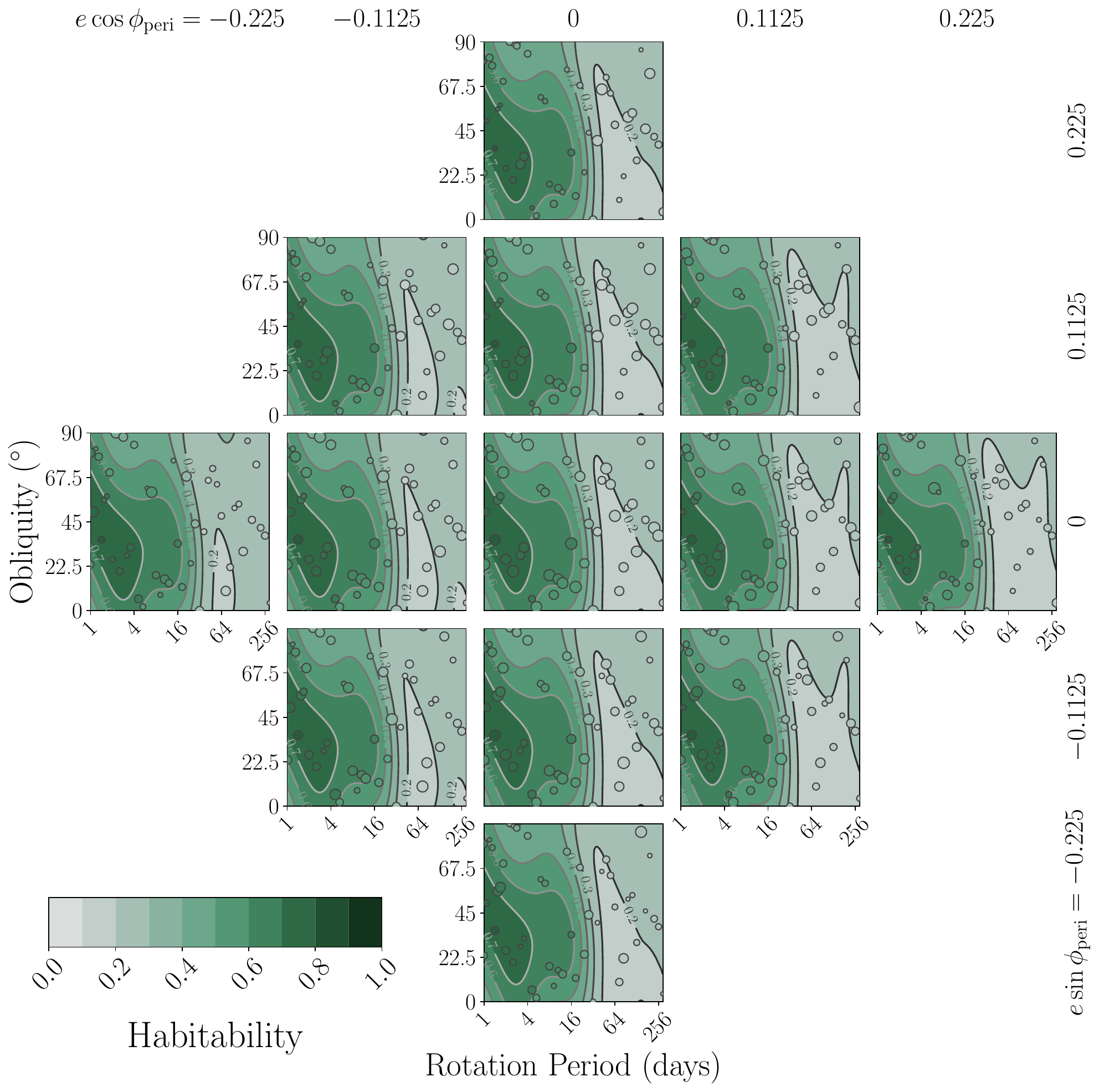} \\
\includegraphics[height=10.5cm]{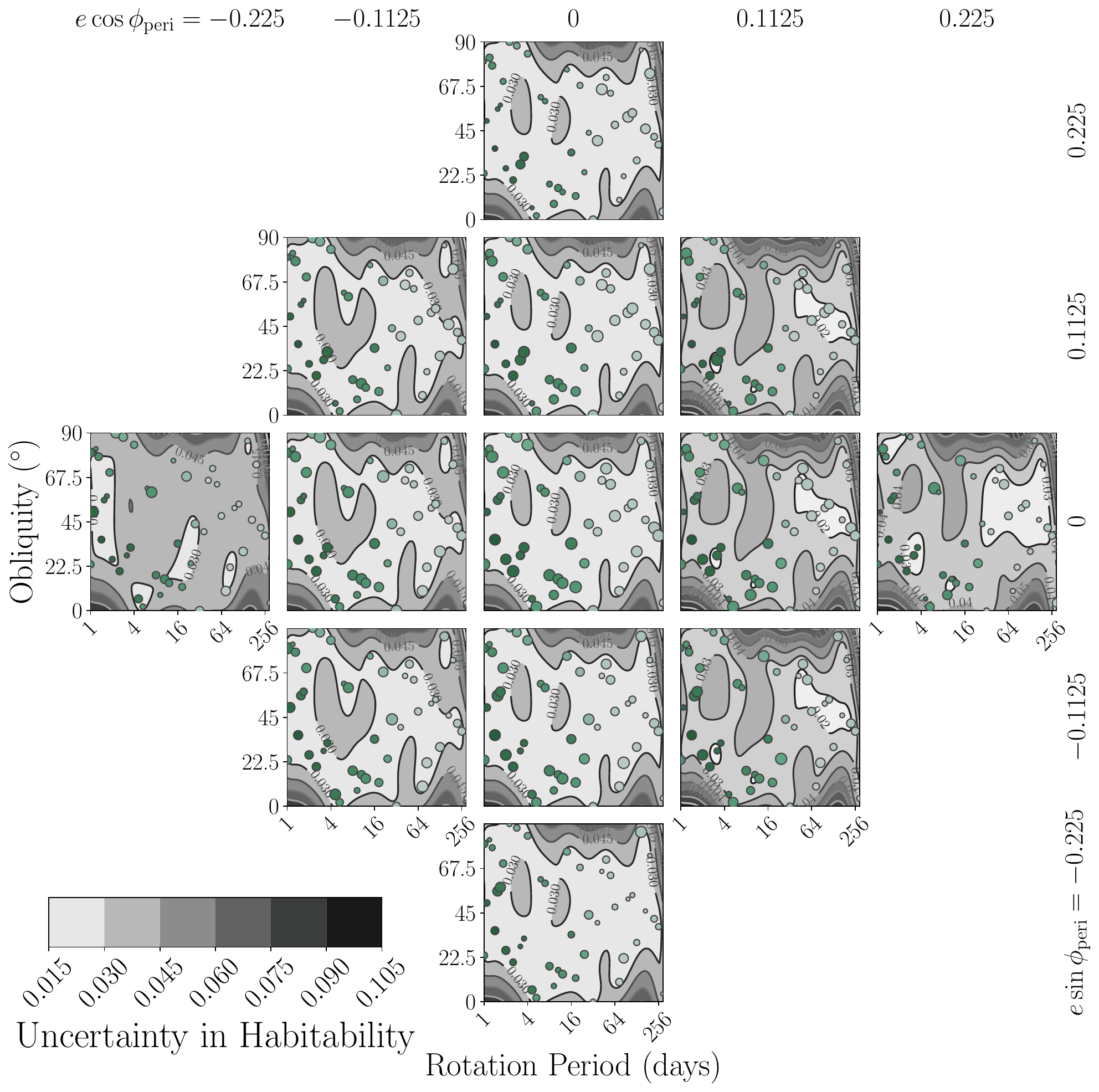} \\
\end{tabular}
\caption{The emulated fractional averages of the habitability metric for the training set. The emulation is shown across the four model dimensions as a ``grid of grids'', where each sub-plot shows rotation period versus obliquity, and the sub-plots are arranged such that eccentricities increase radially from the center sub-plot. The outcomes of the actual runs are over-plotted at their locations as points, colored on the same scale. Each sub-plot contains all model points, but since each sub-plot is fixed at specific values of eccentricity and longitude of periapse, the points' apparent sizes are scaled to represent their ``distance'' in eccentricity from their projected position on that sub-plot. Below the emulated grid we show the corresponding uncertainties in the predictions as grey-scale filled contours, with higher uncertainties as darker shades.}
\label{fig:HabitabilityMetric_training_grid-of-grids}
\end{center}
\end{figure}

Combining these into the habitability metric, it is not surprising that the interpolation follows the trends of the temperature and precipitation: the strongest gradient remains in rotation period, and there is a shallower dip in habitability as obliquity increases at short rotation periods. Broadly speaking, fast rotation periods with low to intermediate obliquities maximize habitability. Reflecting the influence of the temperature metric, the emulator places a maximum habitability of $\approx 0.80$ at a rotation period of 1.41 days and an obliquity of $36^\circ$, with a plateau of habitability values at or above 0.6 for rotation periods $\lesssim 16$ days and at obliquities between roughly 10 and $60^\circ$. This is similar in structure to the broad region of maximum habitability seen for the non-eccentric emulation (right panel of Figure \ref{fig:He-grid_emulation}).

\subsection{Testing the Emulator}\label{sec:results:test}
\begin{figure}[htb!]
\begin{center}
\includegraphics[width=17cm]{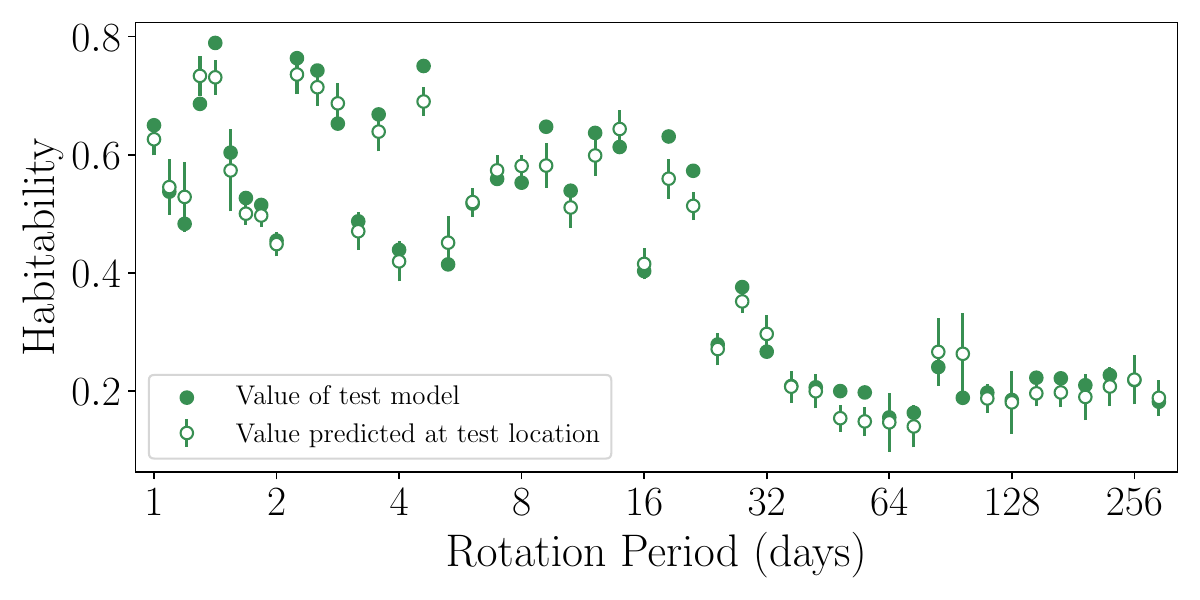}
\caption{A comparison of the habitability metrics from the test models (filled circles), compared with the habitability metric values predicted by the emulator at the locations of the test points (open circles). The error bars are the emulator's estimated uncertainties in its predictions. We order the points by rotation period here to compare along a single dimension.}
\label{fig:emulator_training-vs-test}
\end{center}
\end{figure}

We now introduce the outcomes of the test models, which are generated from a distinct Latin Hypercube sampling of the same parameter space. The goal is to compare the habitability metrics as calculated directly from the test models, with the metric values that the emulator predicts from the training set at the locations of the test points in the 4-dimensional parameter space. Since the emulation process also estimates the uncertainty in its predicted values, we can calculate the residuals in the fits to the test point habitabilities, as shown in Figure \ref{fig:emulator_training-vs-test}. There are 14 points that differ from their emulator-predicted value by more than the emulator's estimated uncertainty (the $1-\sigma$ confidence interval), with 11 of the 14 being underestimates and all occurring at rotation periods shorter than 64 days. 4 points differ by at least 2 $\sigma$, with the two worst cases being Test Cases 16 ($P_\mathrm{rot} = 4.59$ days, $\psi = 22^\circ$, $e=0.055$, $\phi_\mathrm{p}=152^\circ$) and 27 (($P_\mathrm{rot} = 21.1$ days, $\psi = 28^\circ$, $e=0.095$, $\phi_\mathrm{p}=232^\circ$)). This is consistent with the expectation of a normally-distributed set of measurements, which indicates the uncertainty estimates are indeed accurate.

\begin{figure}[htb!]
\begin{center}
\includegraphics[width=8.5cm]{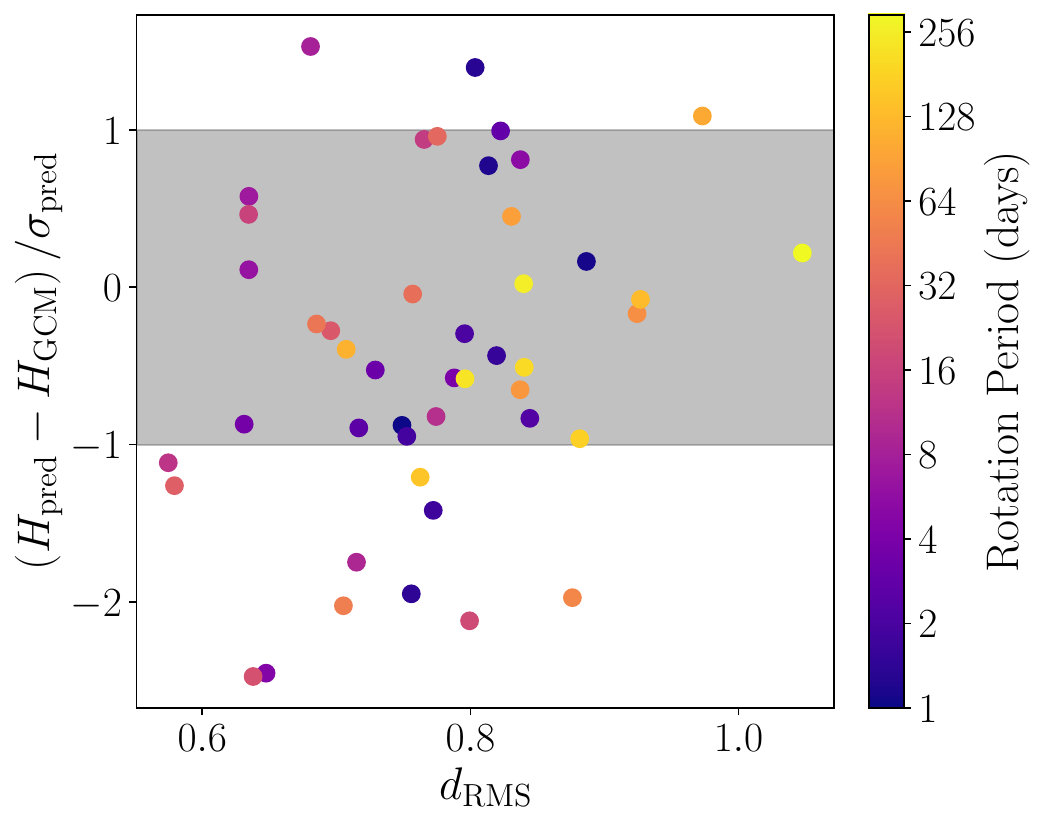}
\caption{The residuals in the emulator predictions of the habitability values from the test set of GCM runs, plotted versus the RMS distance in the parameter space (as defined in Equation \ref{eq:parameter_RMS_distance}). Each point is colored on a log scale by the rotation period of the run. While we see a trend in that the longest rotation period cases tend to lie close to their predicted values, there is no obvious correlation between how well a habitability metric is predicted and either rotation period or RMS distance.}
\label{fig:emulator_test_residuals}
\end{center}
\end{figure}

We explore the hypothesis that the primary factor in determining whether the emulator will be inaccurate in predicting habitability is how densely the training data are in the areas of parameter space being tested. Figure \ref{fig:emulator_test_residuals} shows the test points colored by their ``RMS distance'' in the 4-D space, defined here as
\begin{equation}\label{eq:parameter_RMS_distance}
    d_{i, \mathrm{RMS}} \equiv \left\{ \frac{1}{N_\mathrm{train}} \sum\limits_{j=1}^{N_\mathrm{train}} \left[\frac{\log_2\!\left(P_i/P_j\right)}{\Delta \log_2\!P}\right]^2 + \left(\frac{\psi_i-\psi_j}{\Delta \psi}\right)^2 + \left(\frac{e_i\cos\phi_{\mathrm{p}, i} - e_j\cos\phi_{\mathrm{p}, j}}{\Delta e\cos\phi_{\mathrm{p}}}\right)^2 + \left(\frac{e_i\sin\phi_{\mathrm{p}, i} - e_j\sin\phi_{\mathrm{p}, j}}{{\Delta e\sin_{\mathrm{p}}}}\right)^2 \right\}^{1/2}
\end{equation}
where we have trained the emulator using the logarithm base-2 of the rotation periods and the recasting of eccentricity and longitude. We also normalize by the range of values along each dimension, as represented by the $\Delta$ quantities in the denominators (e.g. $\Delta\psi \equiv \psi_\mathrm{max}-\psi_\mathrm{min} = 90^\circ$). Little correlation exists between the RMS distances in the parameter space and the outcome of the predictions versus the directly modeled habitability values. This does not necessarily imply that the density of sampling is not a factor contributing to outliers, but that the greater contributor may be complex physical behavior in the climate models, rather than statistical effects from the emulation.

\subsection{Predicting Habitability across all Models}\label{sec:results:all-models-habitabiltity}
\begin{figure}[htb!]
\begin{center}
\begin{tabular}{c}
\includegraphics[height=10.5cm]{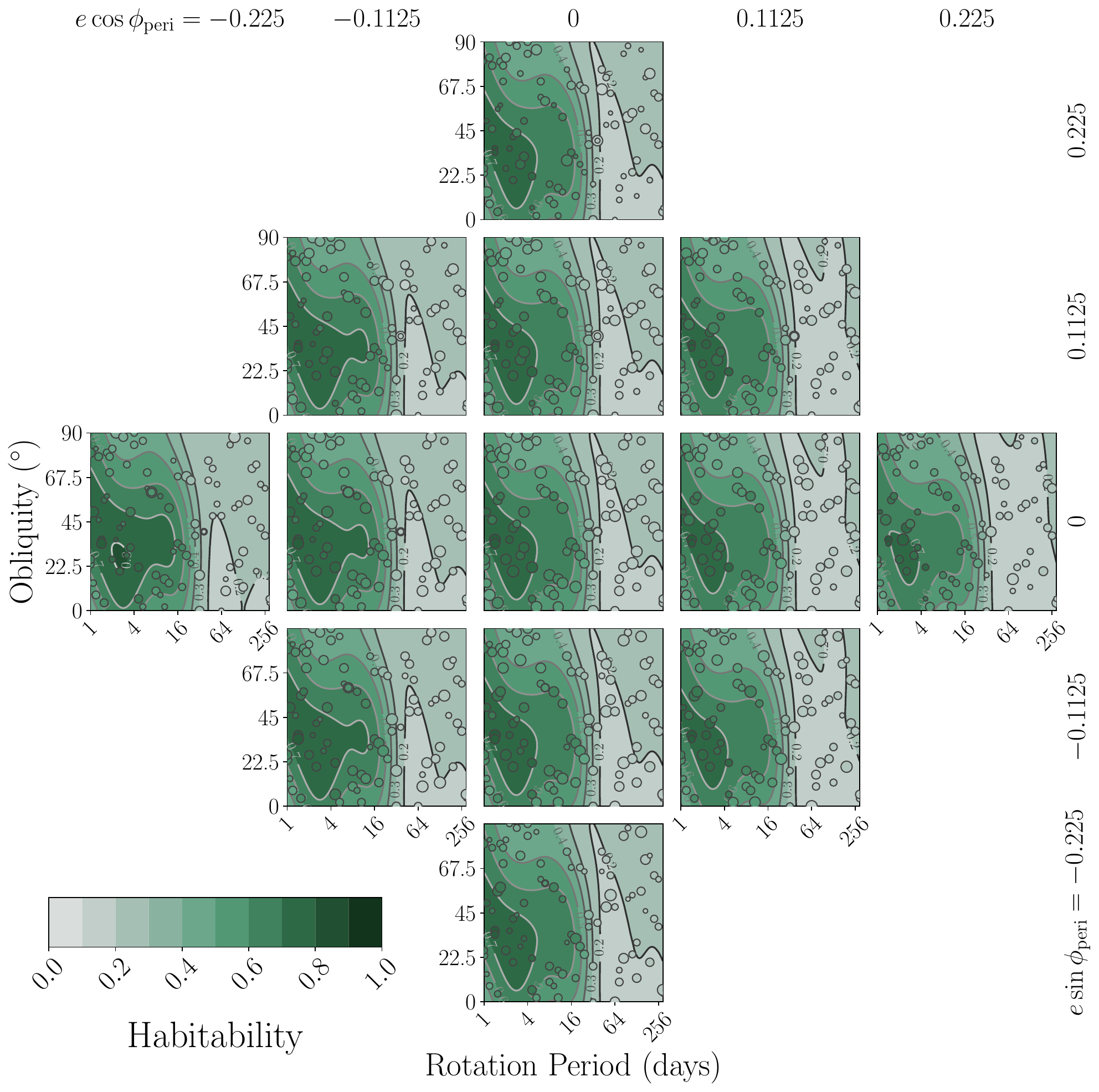} \\
\includegraphics[height=10.5cm]{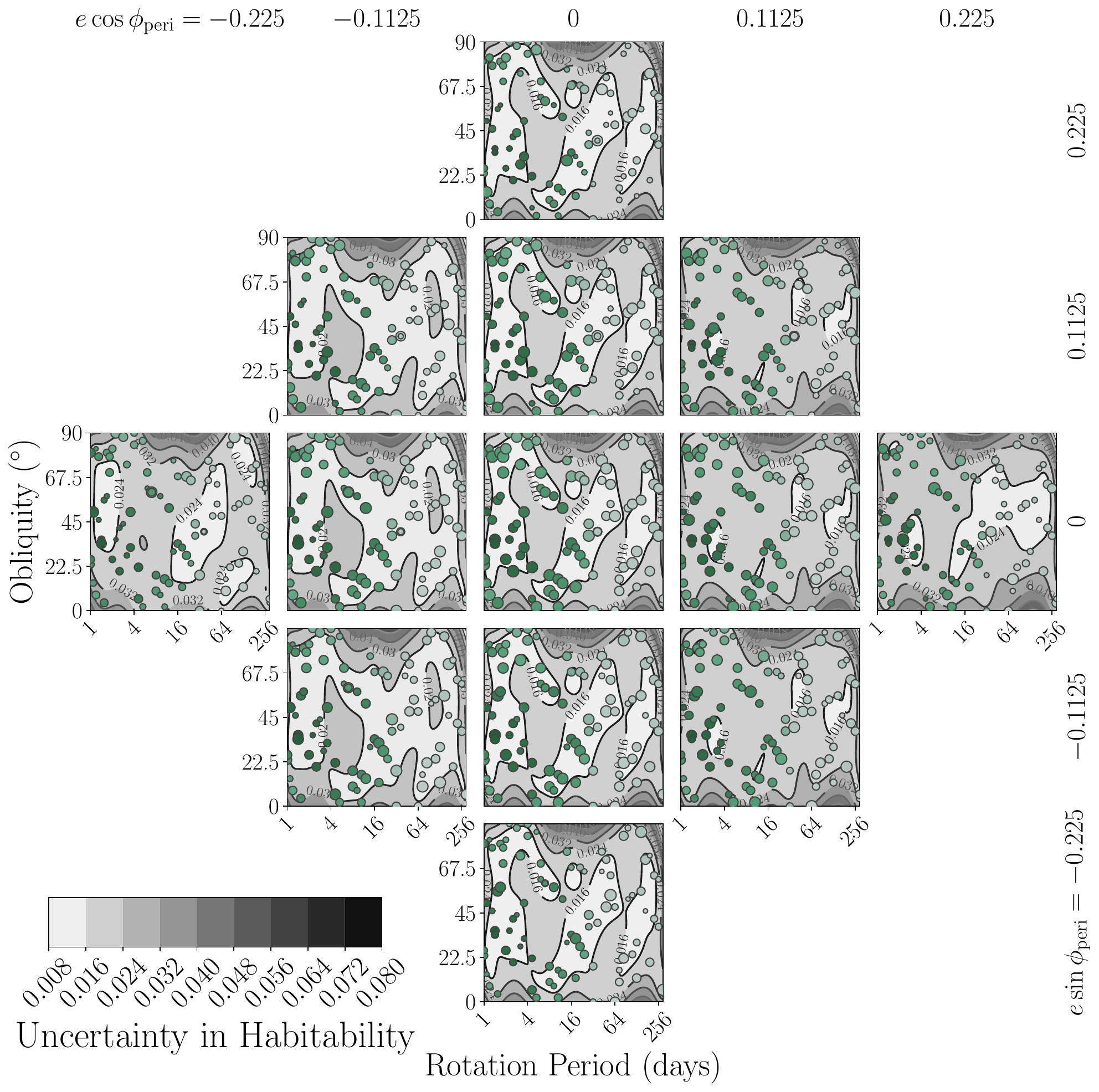} \\
\end{tabular}
\caption{The habitability metric for the combined training and test models (top) and the associated statistical uncertainties in the emulation (bottom). The emulation is shown across the four model dimensions as a ``grid of grids'', where each sub-plot shows rotation period versus obliquity, and the sub-plots are arranged such that eccentricities increase radially from the center sub-plot. The outcomes of the actual runs are over-plotted at their locations as points, colored on the same scale. Each sub-plot contains all model points, but since each sub-plot is fixed at specific values of eccentricity and longitude of periapse, the points' apparent sizes are scaled to represent their ``distance'' in eccentricity from their projected position on that sub-plot.}
\label{fig:HabitabilityMetric_all_grid-of-grids}
\end{center}
\end{figure}

We can use emulation on the combined training and test models, to generate the fullest interpolation of the parameter space (Figure \ref{fig:HabitabilityMetric_all_grid-of-grids}). With all the eccentric models included, we see that the most consistent structure that emerges is that habitability is highest for obliquities up to $\approx 50^\circ$, and rotation periods shorter than $\approx 20$ days. No models with rotation periods longer than 32 days exhibit an average habitability above 0.25, regardless of the obliquity or eccentricity, and overall habitability drops (though to a smaller extent) at faster rotations when comparing high obliquities with low obliquities. Beyond this, it is difficult to determine whether the finer structures seen in the various eccentricity ``slices'' of Figure \ref{fig:HabitabilityMetric_all_grid-of-grids} are robust given the sample size of our climate models. The darkest contours of habitability ($\geq 70$\%) are broadest in the slice at high eccentricity and $\phi_\mathrm{p}$ close to $180^\circ$ (the left-most sub-plot). However, it is difficult to draw any conclusions about whether the observed variations in habitability contours with longitude of periapse in the emulator at the highest eccentricities are more than artifacts of a sparse sampling.

\section{Discussion}\label{sec:discussion}
\subsection{Are Monthly Outputs Sufficient for Calculating Habitability?}\label{sec:discussion:output-timescale}
With our new calculation method outlined in \S \ref{sec:habitability_models:metric}, the fractional temperature habitability shows a significant increase (by a factor of $\sim 2$) compared to the results from \citet{He2022} for the slowest rotation period. As discussed in \S \ref{sec:pre-results:rotation-only}, this change is due to the influence of the diurnal cycle. Our ROCKE-3D simulations create output data at monthly timescales, which are roughly each 1/12 of the orbital period (or about 30 modern Earth days). This limits our ability to characterize the diurnal cycle-induced temperature variations to only the slowest rotation periods (64 days and above), and does imply that the faster rotation periods would also have different fractional habitabilities if the calculation was made from shorter averaging periods. While our monthly simulation output is a decision primarily made for computational practicality, we believe that the timescales of roughly 1 month are appropriate for thinking about a ``growing season'' in the context of habitability. In other words, if the period above 0$^\circ$C driven by diurnal variability is only a few days long, this would not be a long enough ``growing season'' to promote a diverse biosphere with large organisms like trees. Thus, computing the habitability at 1-month timescales does have reasonable physical justifications in addition to the practical issues.

\subsection{A ``Break'' in Habitability at Day Lengths above 20 Earth Days}\label{sec:discussion:rotation-break}
The results of the emulation, as shown in Figure \ref{fig:emulator_training-vs-test}, suggest that there are two fundamental regimes of behavior, largely delineated by the aforementioned division between ``short'' and ``long'' rotation periods. There is a marked decrease in habitability at rotation periods longer than $\sim 20$ days and especially longer than 32 days. At these slow rotations, there is relatively little dependence of habitability on any of the other parameters. At rotation periods shorter than $\sim 20$ days, obliquity plays a role in shaping the habitability structure, primarily through temperature. This effect is known from previous studies and is largely driven by how larger obliquities change the distribution of insolation. Variations in the exhibit finer structures when rotation is faster. Two of these physical drives that have a major, global effect on the climate conditions include variations in the Coriolis force --- which drives atmospheric dynamics such as Hadley circulations --- and obliquity-driven distribution of insolation across the planetary surface, are two key physical drives of this behavior.

\subsection{A Lack of a Significant Eccentricity Dependence in the Land and Annual Mean}\label{sec:discussion:eccentricity}
Given that even small values of eccentricity can induce remarkable changes to the temperature and precipitation profile, as inferred to have occurred through Earth's history, we might expect a significant signal imparted in the emulated habitability landscape. However, because our habitability metric as a single value averages over land area and time, this signal appears to wash out in the means. Recall that, because each semi-major axis is scaled to keep fixed the total instellation over one orbit, this statement holds in isolation of other factors influencing the instellation. In the case of long rotation periods and/or high obliquities, we see reductions principally because there are extended periods of time where large fractions of the planet linger in darkness. In the case of eccentricity-obliquity interactions, we do not change the day-night cycles themselves, but rather how intense the daytime insolation will be over the course of the year. A summer with 60\% the typical solar heating, as could occur for one hemisphere at apoapse in the most extreme eccentricity case we have modeled, would certainly cool the affected hemisphere, but the effect may not be so extreme as to render it uninhabitable by temperature alone. The real-life example of the humid Sahara is primarily a precipitation effect, which certainly could have an effect on our climate habitability. However, such a shift would need to be taken in a global context; as northern deserts get wetter, so too could regions at similar latitudes in the southern hemisphere dry out due to a corresponding decrease in their summertime insolation. This would depend on the relative land areas in each hemisphere in a complex way, but asymmetries may very well only impart an effect that is too small to be captured in our emulation over the means.

\subsection{Rotation Rate Constraints will be Key for Applying Habitability Studies to Observations}\label{sec:discussion:prescriptions}
This work advances evidence that, given an annual instellation, surface topography, and ocean cover similar to Earth's, the rotation period will remain the primary influence on average temperature and precipitation-based habitability --- even with significant orbital eccentricity. Wide parameter studies such as this will yield more refined model studies as we unearth the underlying true distribution of rotation states for terrestrial exoplanets; therefore, the priority for accurate rotation constraints will be crucial. While observational constraints on orbital elements for small planets are improving with the rise of the current generation of high precision radial velocity measurements \citep[see e.g.][]{VanEylen2021,Passegger2024,Brady2024}, constraining rotation is more difficult, particularly at orbital distances beyond where planets are likely to be tidally locked to their host stars. With the next generation of direct imaging observatories, it may be possible to constrain rotation periods with time series observations at sufficient signal-to-noise to detect broadband or spectral variations.

\section{Conclusions}\label{sec:conclusion}
We present the results of applying modifications to existing habitability metrics to construct a statistical emulator of habitability across rotation period, obliquity, and orbital eccentricity. A set of ROCKE-3D climate models was constructed to span a range of these spin and orbit parameters, using a Latin Hypercube Sampling approach to distribute them within the parameter space. We find that:
\begin{itemize}
    \item Our comparison of habitability metrics from previous works shows that, in order to accurately model the metrics at the slowest rotation periods, intermediate averaging should not be done on sub-year model outputs when the rotation period of the model exceeds the model time output interval.
    \item Emulation on the model grid of \citet{He2022} reproduces the structure seen in rotation period and obliquity, with local maxima in temperate area at rotation periods of 1 Earth day and intermediate obliquities ($\psi \approx 50^\circ$), as well as at rotation periods $\sim 16$ days with low obliquity.
    \item While there is additional structure seen in our emulated landscape of habitability as a function of orbital eccentricity, its effect is small when compared with the effects of rotation period and obliquity, and is not significant enough to make any robust conclusions about the effect of eccentricity on habitability.
\end{itemize}
As we learn more about not just rotation states but, for example, the possible continental arrangements on Earth-similar worlds, we will be able to apply these statistical techniques to incrementally refine our understanding of the landscapes of habitability for terrestrial worlds. While the emulator is not perfect in reproducing the habitabilities of the test cases, it shows that the methodology is a plausible approach to illuminating the possible landscapes of habitability. Future steps include deeper dives into the spatial and time variability of habitability, refinements to the habitability metric, and exploring the role that Earth's continental motions have had on the global climate.

\software{Astropy \citep{ast13}, Jupyter \citep{klu16}, Matplotlib \citep{hun07}, Numpy \citep{van11}, Pandas \citep{mck10}, pyDOE \citep{pyDOE}, Scikit-learn \citep{scikit-learn}, Scipy \citep{jon01}, xarray \citep{xarray}. \\ \\ The Python code and scripts used to run the emulation are available in a Github repository at \url{https://github.com/adadams/habitability_metrics}, archived at \citet{AdamsHM2024}.}

\appendix

\section{Temperature and Precipitation Habitability Grids with All Models}\label{sec:appendix:all-model-grids}
\begin{figure}[htb!]
\begin{center}
\begin{tabular}{c}
\includegraphics[height=10.5cm]{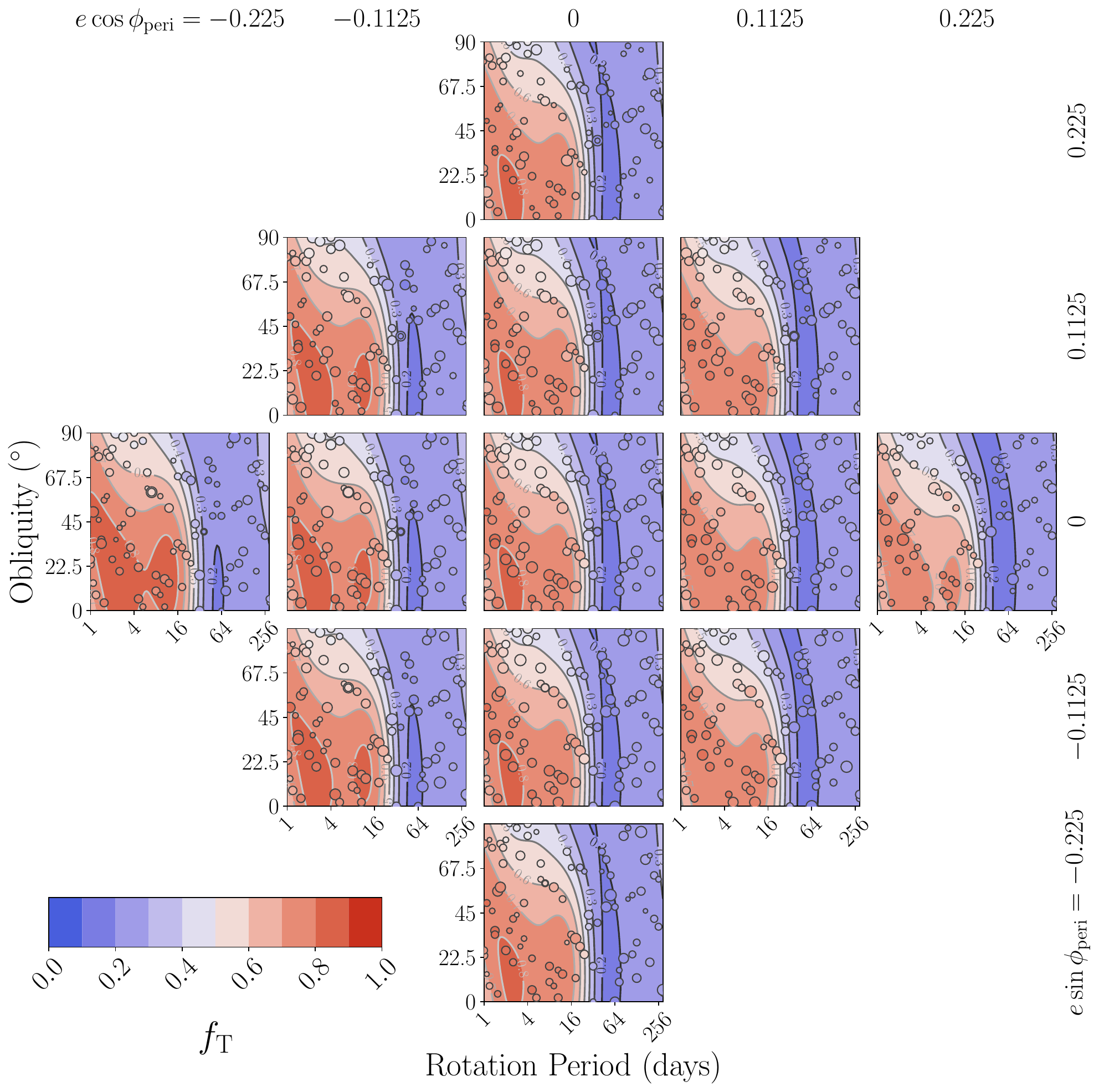} \\
\includegraphics[height=10.5cm]{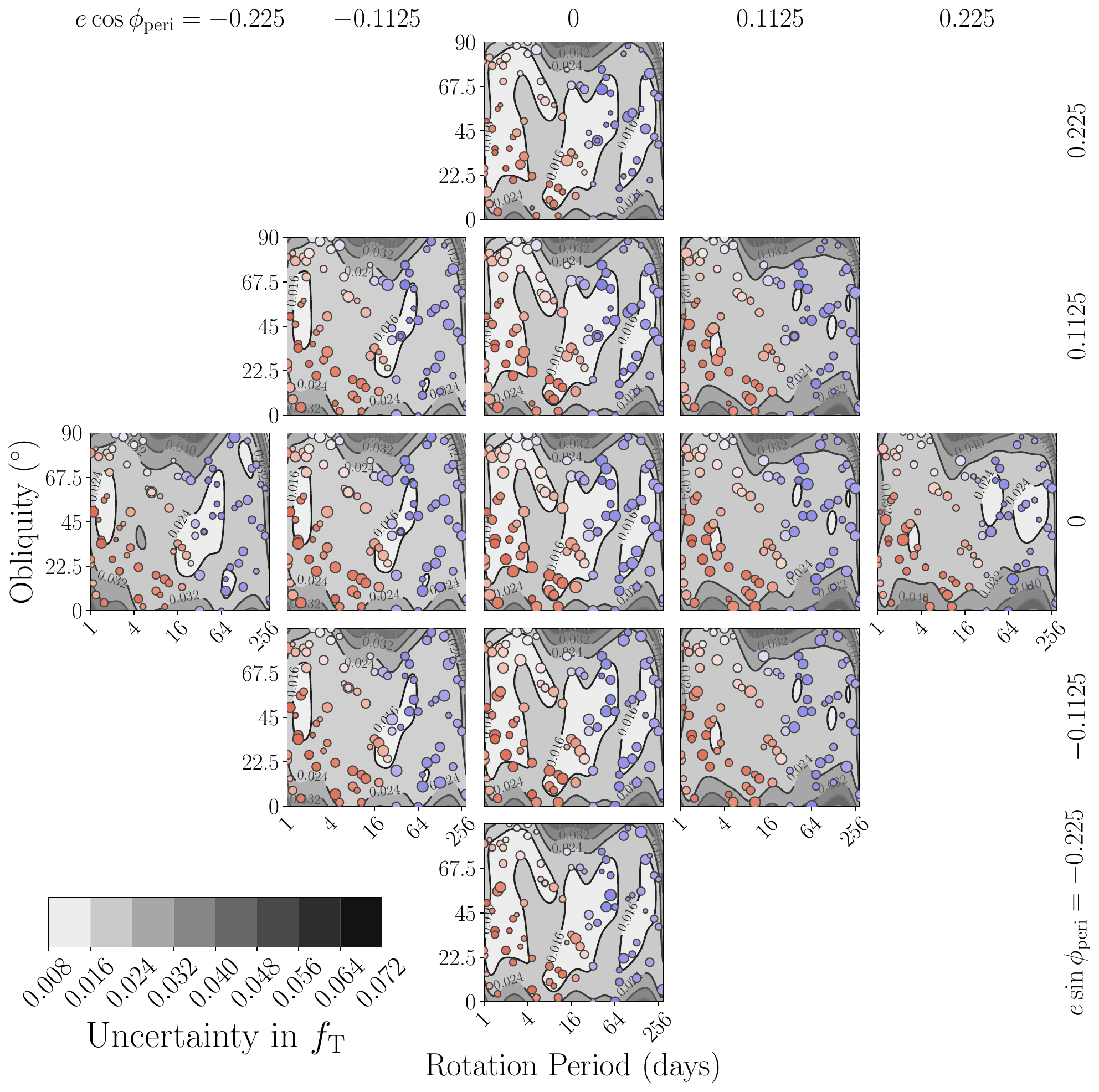} \\
\end{tabular}
\caption{The temperature metric ($f_\mathrm{T}$) for the combined training and test models (top) and the associated statistical uncertainties in the emulation (bottom). The emulation is shown across the four model dimensions as a ``grid of grids'', where each sub-plot shows rotation period versus obliquity, and the sub-plots are arranged such that eccentricities increase radially from the center sub-plot. The outcomes of the actual runs are over-plotted at their locations as points, colored on the same scale. Each sub-plot contains all model points, but since each sub-plot is fixed at specific values of eccentricity and longitude of periapse, the points' apparent sizes are scaled to represent their ``distance'' in eccentricity from their projected position on that sub-plot.}
\label{fig:TemperatureMetric_all_grid-of-grids}
\end{center}
\end{figure}

\begin{figure}[htb!]
\begin{center}
\begin{tabular}{c}
\includegraphics[height=10.5cm]{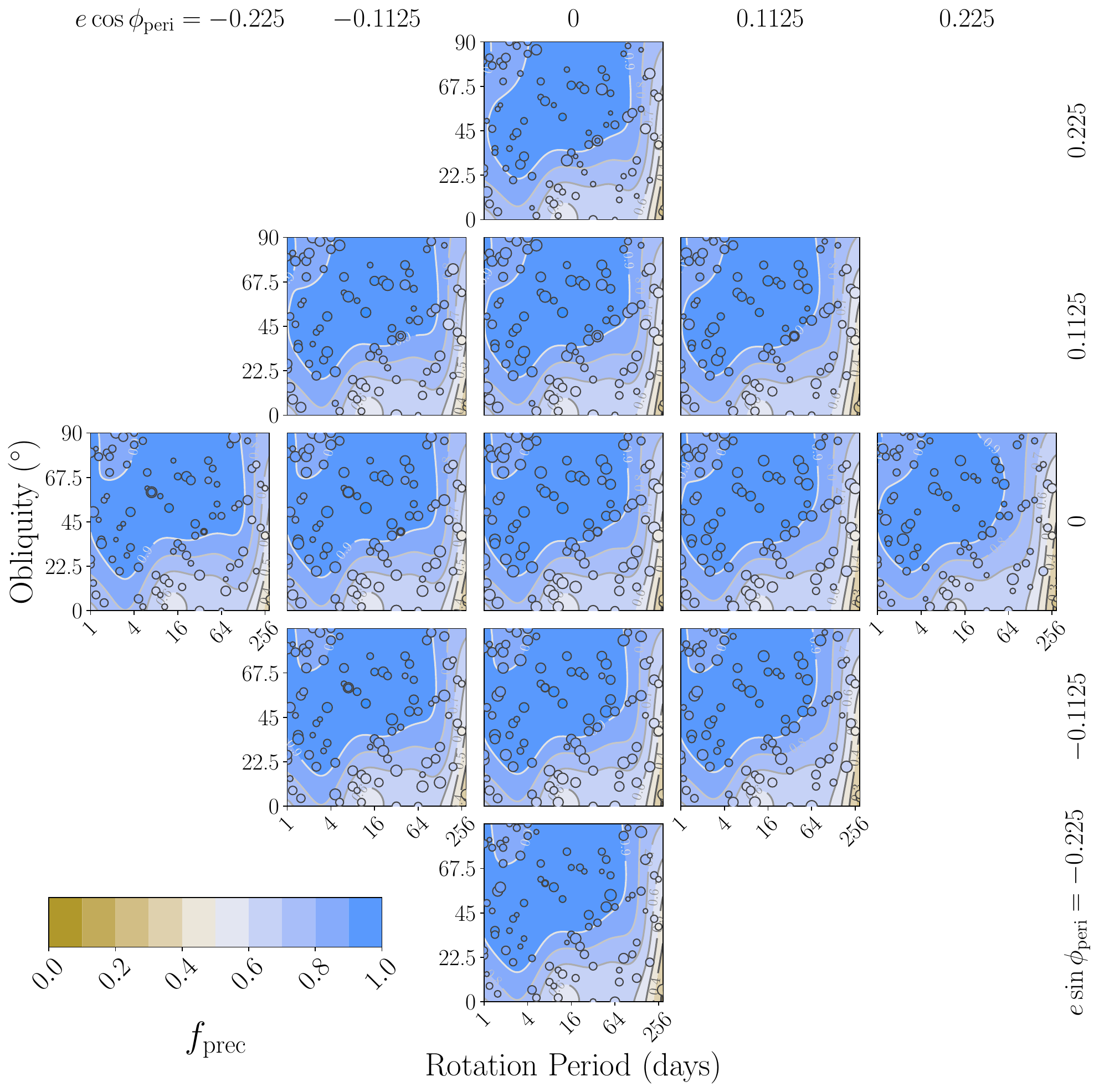} \\
\includegraphics[height=10.5cm]{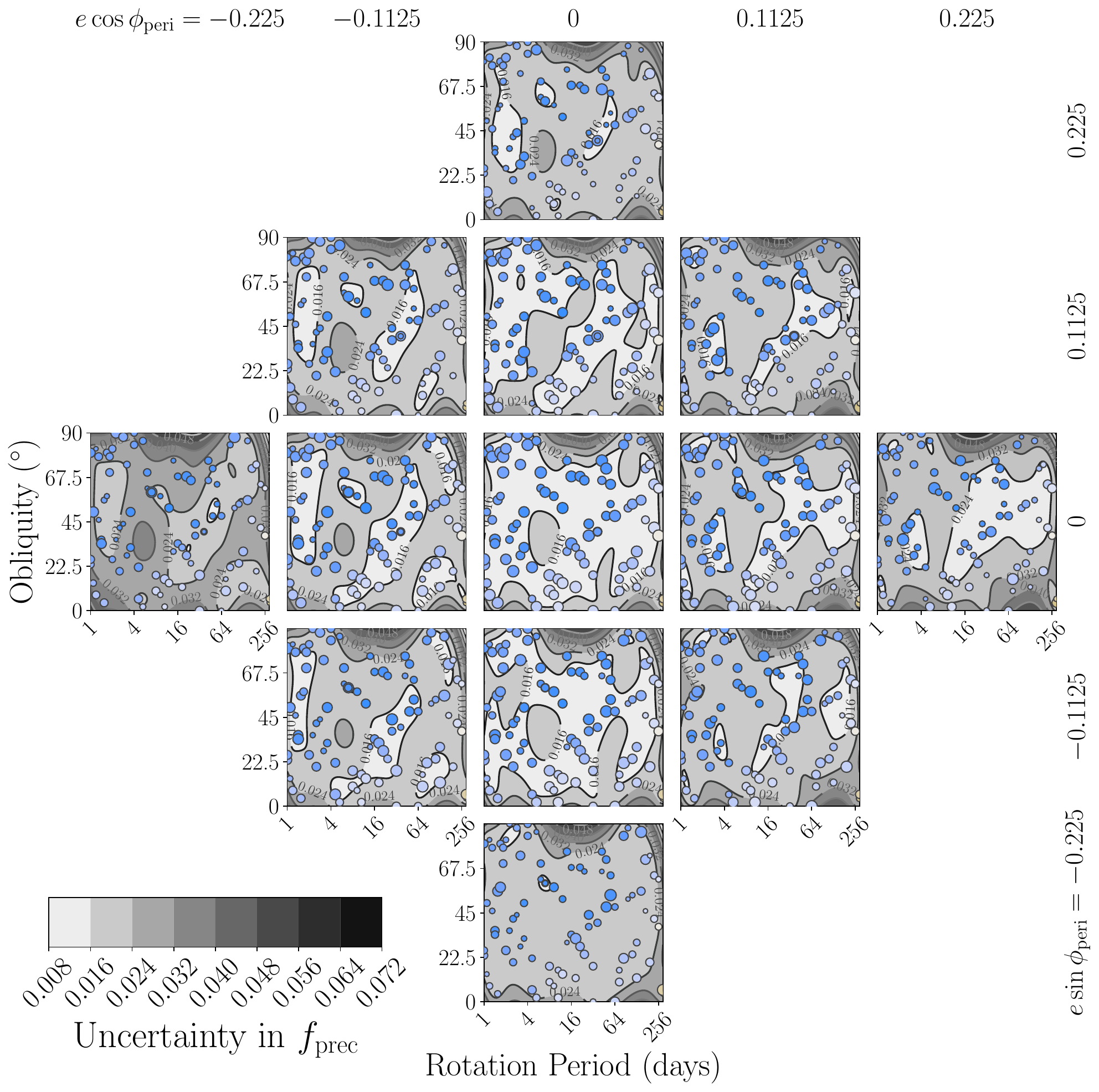} \\
\end{tabular}
\caption{The precipitation metric ($f_\mathrm{prec}$) for the combined training and test models (top) and the associated statistical uncertainties in the emulation (bottom). The emulation is shown across the four model dimensions as a ``grid of grids'', where each sub-plot shows rotation period versus obliquity, and the sub-plots are arranged such that eccentricities increase radially from the center sub-plot. The outcomes of the actual runs are over-plotted at their locations as points, colored on the same scale. Each sub-plot contains all model points, but since each sub-plot is fixed at specific values of eccentricity and longitude of periapse, the points' apparent sizes are scaled to represent their ``distance'' in eccentricity from their projected position on that sub-plot.}
\label{fig:PrecipitationMetric_all_grid-of-grids}
\end{center}
\end{figure}

\section{Outcomes of Emulator Tests with Varying Temperature and Precipitation Thresholds}\label{sec:appendix:metric-threshold-sensitivity}

\begin{figure}[htb!]
\begin{center}
\begin{tabular}{cc}
$0 < T_\mathrm{surf} < 50^\circ$ C & $-20 < T_\mathrm{surf} < 100^\circ$ C \\
\includegraphics[width=8.5cm]{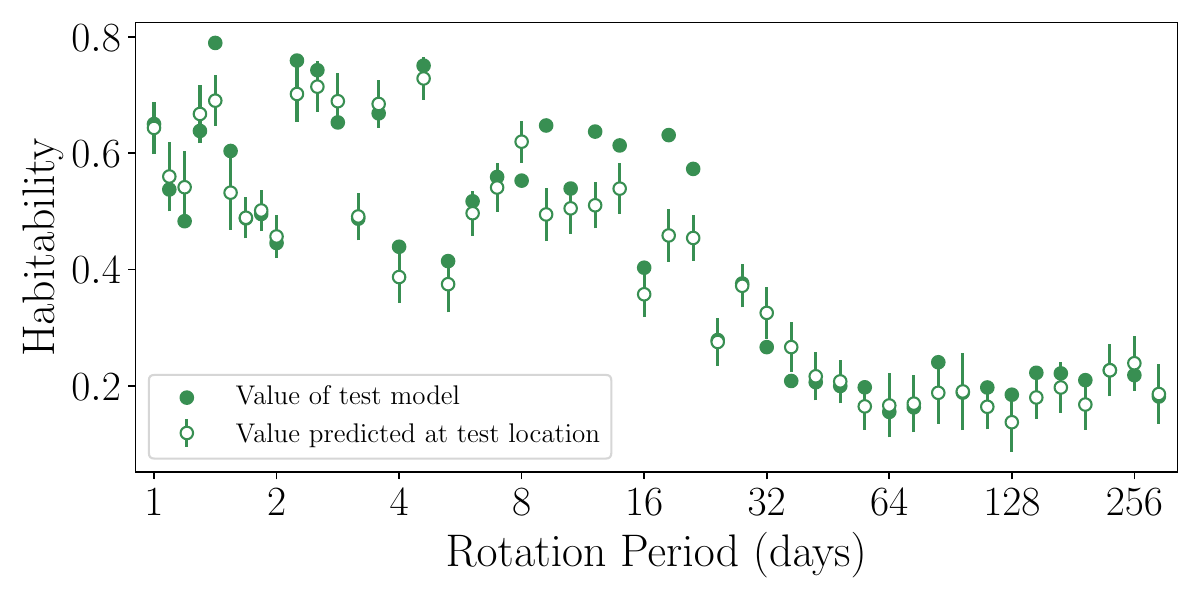} &
\includegraphics[width=8.5cm]{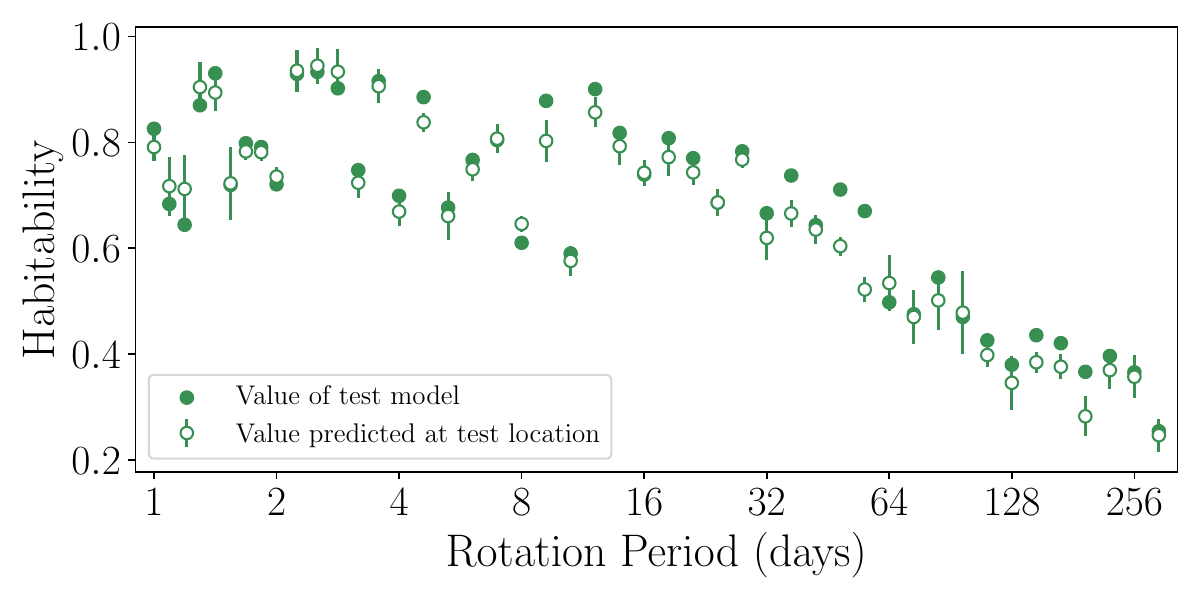} \\
Precipitation $ > 600$ mm & Precipitation $ > 150$ mm \\
\includegraphics[width=8.5cm]{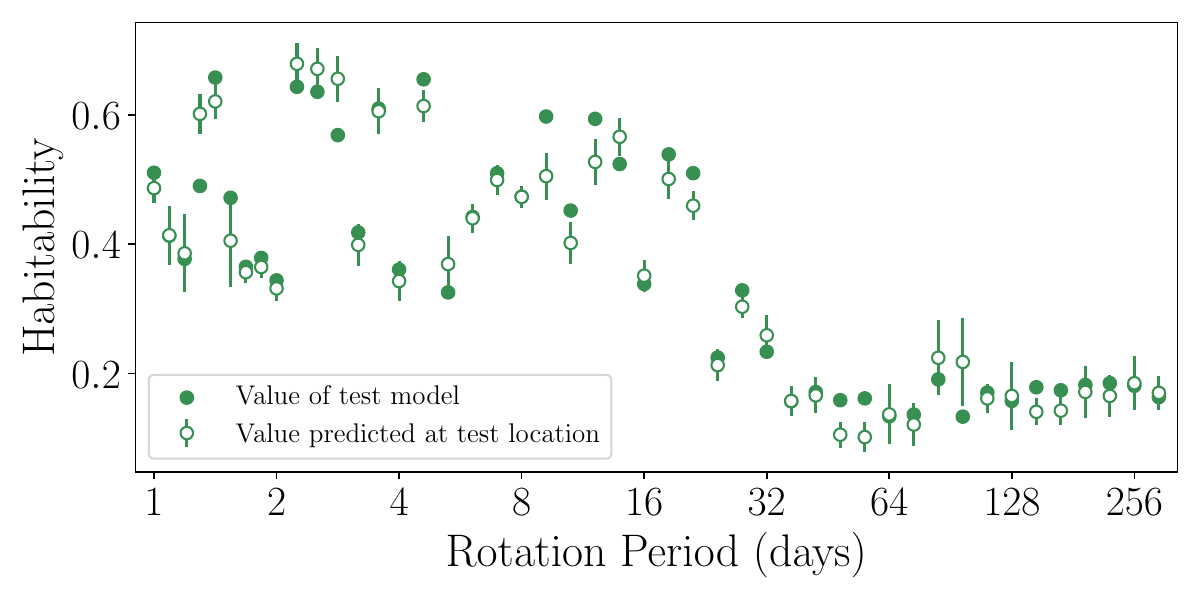} &
\includegraphics[width=8.5cm]{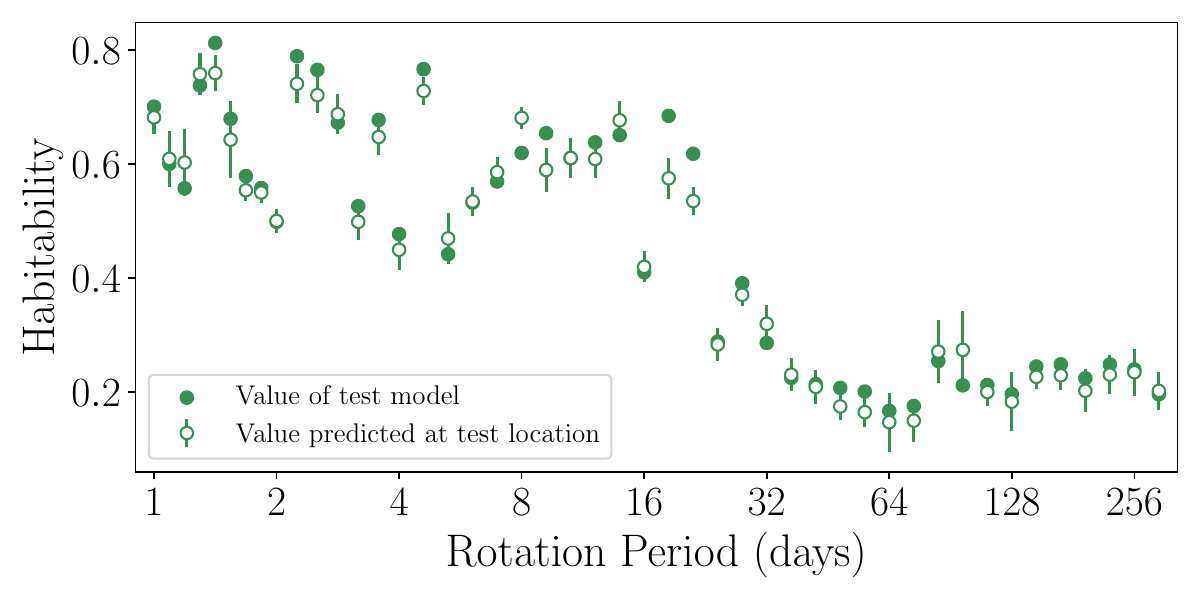} \\
\end{tabular}
\caption{Similar to Figure \ref{fig:emulator_training-vs-test}, varying the thresholds of the temperature and precipitation. Each subplot is a comparison of the habitability metrics from the test models (filled circles), compared with the habitability metric values predicted by the emulator at the locations of the test points (open circles). The error bars are the emulator's estimated uncertainties in its predictions. We order the points by rotation period here to compare along a single dimension.}
\label{fig:emulator_training-vs-test_varying-thresholds}
\end{center}
\end{figure}

\begin{acknowledgements}
This work was funded by a Habitable Worlds grant through the NASA Research Opportunities in Space and Earth Sciences (ROSES) program (PI: Margaret Turnbull, award number 80NSSC21K1703).
\end{acknowledgements}

\bibliography{library}{}

\begin{thebibliography}{}
\expandafter\ifx\csname natexlab\endcsname\relax\def\natexlab#1{#1}\fi
\providecommand{\url}[1]{\href{#1}{#1}}
\providecommand{\dodoi}[1]{doi:~\href{http://doi.org/#1}{\nolinkurl{#1}}}
\providecommand{\doeprint}[1]{\href{http://ascl.net/#1}{\nolinkurl{http://ascl.net/#1}}}
\providecommand{\doarXiv}[1]{\href{https://arxiv.org/abs/#1}{\nolinkurl{https://arxiv.org/abs/#1}}}

\bibitem[{Abe {et~al.}(2011)Abe, Abe-Ouchi, Sleep, \& Zahnle}]{Abe2011}
Abe, Y., Abe-Ouchi, A., Sleep, N.~H., \& Zahnle, K.~J. 2011, Astrobiology, 11, 443, \dodoi{10.1089/ast.2010.0545}

\bibitem[{Adams(2024)}]{AdamsHM2024}
Adams, A. 2024, adadams/habitability\_metrics: v1.0.1,  Zenodo, \dodoi{10.5281/zenodo.14556043}

\bibitem[{An {et~al.}(2023)An, Xie, Dai, \& Zhou}]{An2023}
An, D.-S., Xie, J.-W., Dai, Y.-Z., \& Zhou, J.-L. 2023, The Astronomical Journal, 165, 125, \dodoi{10.3847/1538-3881/acb533}

\bibitem[{Armstrong {et~al.}(2014)Armstrong, Barnes, Domagal-Goldman, Breiner, Quinn, \& Meadows}]{Armstrong2014}
Armstrong, J., Barnes, R., Domagal-Goldman, S., {et~al.} 2014, Astrobiology, 14, 277, \dodoi{10.1089/ast.2013.1129}

\bibitem[{Arnold {et~al.}(2009)Arnold, Bréon, \& Brewer}]{Arnold2009}
Arnold, L., Bréon, F.-M., \& Brewer, S. 2009, International Journal of Astrobiology, 8, 81, \dodoi{10.1017/S1473550409004406}

\bibitem[{{Astropy Collaboration} {et~al.}(2013){Astropy Collaboration}, Robitaille, Tollerud, Greenfield, Droettboom, Bray, Aldcroft, Davis, Ginsburg, Price-Whelan, Kerzendorf, Conley, Crighton, Barbary, Muna, Ferguson, Grollier, Parikh, Nair, Unther, Deil, Woillez, Conseil, Kramer, Turner, Singer, Fox, Weaver, Zabalza, Edwards, Azalee~Bostroem, Burke, Casey, Crawford, Dencheva, Ely, Jenness, Labrie, Lim, Pierfederici, Pontzen, Ptak, Refsdal, Servillat, \& Streicher}]{ast13}
{Astropy Collaboration}, Robitaille, T., Tollerud, E., {et~al.} 2013, {\textbackslash}aap, 558, A33, \dodoi{10.1051/0004-6361/201322068}

\bibitem[{Bar-On {et~al.}(2018)Bar-On, Phillips, \& Milo}]{Bar-On2018}
Bar-On, Y.~M., Phillips, R., \& Milo, R. 2018, Proceedings of the National Academy of Sciences, 115, 6506, \dodoi{10.1073/pnas.1711842115}

\bibitem[{Berger {et~al.}(1993)Berger, Loutre, \& Tricot}]{ber93}
Berger, A., Loutre, M.-F., \& Tricot, C. 1993, {\textbackslash}jgr, 98, 10, \dodoi{10.1029/93JD00222}

\bibitem[{Brady {et~al.}(2024)Brady, Bean, Seifahrt, Kasper, Luque, Stefánsson, Stürmer, Charbonneau, Collins, Doty, Essack, Fukui, Grau~Horta, Hedges, Hellier, Jenkins, Narita, Quinn, Shporer, Schwarz, Seager, Stassun, Striegel, Watkins, Winn, \& Zambelli}]{Brady2024}
Brady, M., Bean, J.~L., Seifahrt, A., {et~al.} 2024, The Astronomical Journal, 168, 67, \dodoi{10.3847/1538-3881/ad500a}

\bibitem[{Colose {et~al.}(2021)Colose, Haqq-Misra, Wolf, Del~Genio, Barnes, Way, \& Ruedy}]{Colose2021}
Colose, C.~M., Haqq-Misra, J., Wolf, E.~T., {et~al.} 2021, The Astrophysical Journal, 921, 25, \dodoi{10.3847/1538-4357/ac135c}

\bibitem[{Dai {et~al.}(2020)Dai, Huang, Rose, Zhu, \& Tian}]{DaiAiguo2020}
Dai, A., Huang, D., Rose, B. E.~J., Zhu, J., \& Tian, X. 2020, Climate Dynamics, 54, 4515, \dodoi{10.1007/s00382-020-05242-1}

\bibitem[{Deitrick {et~al.}(2018)Deitrick, Barnes, Bitz, Fleming, Charnay, Wilhelm, Armstrong, Quinn, \& Meadows}]{Deitrick2018}
Deitrick, R., Barnes, R., Bitz, C., {et~al.} 2018, The Astronomical Journal, 155, 266, \dodoi{10.3847/1538-3881/aac214}

\bibitem[{Dobrovolskis(2013)}]{Dobrovolskis2013}
Dobrovolskis, A.~R. 2013, Icarus, 226, 760, \dodoi{10.1016/j.icarus.2013.06.026}

\bibitem[{Driscoll \& Barnes(2015)}]{Driscoll2015}
Driscoll, P., \& Barnes, R. 2015, Astrobiology, 15, 739, \dodoi{10.1089/ast.2015.1325}

\bibitem[{Driscoll \& Bercovici(2013)}]{Driscoll2013}
Driscoll, P., \& Bercovici, D. 2013, Icarus, 226, 1447, \dodoi{10.1016/j.icarus.2013.07.025}

\bibitem[{Edwards(1996)}]{Edwards1996a}
Edwards, J.~M. 1996, Journal of Atmospheric Sciences, 53, 1921 , \dodoi{10.1175/1520-0469(1996)053<1921:ECOIFA>2.0.CO;2}

\bibitem[{Edwards \& Slingo(1996)}]{Edwards1996b}
Edwards, J.~M., \& Slingo, A. 1996, Quarterly Journal of the Royal Meteorological Society, 122, 689, \dodoi{10.1002/qj.49712253107}

\bibitem[{Fletcher {et~al.}(2022)Fletcher, McNally, Virgin, \& King}]{Fletcher2022}
Fletcher, C.~G., McNally, W., Virgin, J.~G., \& King, F. 2022, Journal of Advances in Modeling Earth Systems, 14, \dodoi{10.1029/2021MS002836}

\bibitem[{Foley \& Driscoll(2016)}]{Foley2016}
Foley, B.~J., \& Driscoll, P.~E. 2016, Geochemistry, Geophysics, Geosystems, 17, 1885, \dodoi{10.1002/2015GC006210}

\bibitem[{Foley \& Smye(2018)}]{Foley2018}
Foley, B.~J., \& Smye, A.~J. 2018, Astrobiology, 18, 873, \dodoi{10.1089/ast.2017.1695}

\bibitem[{Godolt {et~al.}(2016)Godolt, Grenfell, Kitzmann, Kunze, Langematz, Patzer, Rauer, \& Stracke}]{Godolt2016}
Godolt, M., Grenfell, J.~L., Kitzmann, D., {et~al.} 2016, Astronomy \& Astrophysics, 592, A36, \dodoi{10.1051/0004-6361/201628413}

\bibitem[{Gregoire {et~al.}(2011)Gregoire, Valdes, Payne, \& Kahana}]{Gregoire2010}
Gregoire, L.~J., Valdes, P.~J., Payne, A.~J., \& Kahana, R. 2011, Climate Dynamics, 37, 705, \dodoi{10.1007/s00382-010-0934-8}

\bibitem[{Haqq-Misra {et~al.}(2024)Haqq-Misra, Wolf, Fauchez, \& Kopparapu}]{Haqq-Misra2024a}
Haqq-Misra, J., Wolf, E.~T., Fauchez, T.~J., \& Kopparapu, R.~K. 2024, The Planetary Science Journal, 5, 140, \dodoi{10.3847/PSJ/ad50a7}

\bibitem[{He {et~al.}(2022)He, Merrelli, L’Ecuyer, \& Turnbull}]{He2022}
He, F., Merrelli, A., L’Ecuyer, T.~S., \& Turnbull, M.~C. 2022, The Astrophysical Journal, 933, 62, \dodoi{10.3847/1538-4357/ac6951}

\bibitem[{Hill {et~al.}(2023)Hill, Bott, Dalba, Fetherolf, Kane, Kopparapu, Li, \& Ostberg}]{Hill2023}
Hill, M.~L., Bott, K., Dalba, P.~A., {et~al.} 2023, The Astronomical Journal, 165, 34, \dodoi{10.3847/1538-3881/aca1c0}

\bibitem[{Hoyer \& Hamman(2017)}]{xarray}
Hoyer, S., \& Hamman, J. 2017, Journal of Open Research Software, 5, \dodoi{10.5334/jors.148}

\bibitem[{Hunter(2007)}]{hun07}
Hunter, J.~D. 2007, Computing in Science Engineering, 9, 90, \dodoi{10.1109/MCSE.2007.55}

\bibitem[{Jansen {et~al.}(2019)Jansen, Scharf, Way, \& Del~Genio}]{Jansen2019}
Jansen, T., Scharf, C., Way, M., \& Del~Genio, A. 2019, The Astrophysical Journal, 875, 79, \dodoi{10.3847/1538-4357/ab113d}

\bibitem[{Jones {et~al.}(2001)Jones, Oliphant, Peterson, \& {others}}]{jon01}
Jones, E., Oliphant, T., Peterson, P., \& {others}. 2001, \{{SciPy}\}: {Open} source scientific tools for \{{Python}\}.
\newblock \url{http://www.scipy.org/}

\bibitem[{Kane {et~al.}(2012)Kane, Ciardi, Gelino, \& von Braun}]{Kane2012b}
Kane, S.~R., Ciardi, D.~R., Gelino, D.~M., \& von Braun, K. 2012, Monthly Notices of the Royal Astronomical Society, 425, 757, \dodoi{10.1111/j.1365-2966.2012.21627.x}

\bibitem[{Kane \& Torres(2017)}]{Kane2017}
Kane, S.~R., \& Torres, S.~M. 2017, The Astronomical Journal, 154, 204, \dodoi{10.3847/1538-3881/aa8fce}

\bibitem[{Kane {et~al.}(2016)Kane, Hill, Kasting, Kopparapu, Quintana, Barclay, Batalha, Borucki, Ciardi, Haghighipour, Hinkel, Kaltenegger, Selsis, \& Torres}]{Kane2016}
Kane, S.~R., Hill, M.~L., Kasting, J.~F., {et~al.} 2016, The Astrophysical Journal, 830, 1, \dodoi{10.3847/0004-637X/830/1/1}

\bibitem[{Kasting {et~al.}(1993)Kasting, Whitmire, \& Reynolds}]{kas93}
Kasting, J.~F., Whitmire, D.~P., \& Reynolds, R.~T. 1993, Icarus, 101, 108, \dodoi{10.1006/icar.1993.1010}

\bibitem[{Kiang {et~al.}(2021)Kiang, Colose, Ruedy, Barnes, Elsaesser, Harman, Kane, Russell, Lier-Walqui, Wolf, Aleinov, \& Kiang}]{Kiang2021}
Kiang, N.~Y., Colose, C., Ruedy, R., {et~al.} 2021, 53

\bibitem[{Kluyver {et~al.}(2016)Kluyver, Ragan-Kelley, Perez, Granger, Bussonnier, Frederic, Kelley, Hamrick, Grout, Corlay, Ivanov, Avila, Abdalla, Willing, \& Team}]{klu16}
Kluyver, T., Ragan-Kelley, B., Perez, F., {et~al.} 2016, in Positioning and {Power} in {Academic} {Publishing}: {Players}, {Agents} and {Agendas}, ed. F.~Loizides \& B.~Scmidt (IOS Press), 87--90, \dodoi{10.3233/978-1-61499-649-1-87}

\bibitem[{Kopparapu {et~al.}(2014)Kopparapu, Ramirez, SchottelKotte, Kasting, Domagal-Goldman, \& Eymet}]{kop14}
Kopparapu, R., Ramirez, R., SchottelKotte, J., {et~al.} 2014, {\textbackslash}apjl, 787, L29, \dodoi{10.1088/2041-8205/787/2/L29}

\bibitem[{Kopparapu {et~al.}(2013)Kopparapu, Ramirez, Kasting, Eymet, Robinson, Mahadevan, Terrien, Domagal-Goldman, Meadows, \& Deshpande}]{kop13}
Kopparapu, R.~K., Ramirez, R., Kasting, J.~F., {et~al.} 2013, The Astrophysical Journal, 765, 131, \dodoi{10.1088/0004-637X/765/2/131}

\bibitem[{Leconte {et~al.}(2013)Leconte, Forget, Charnay, Wordsworth, Selsis, Millour, \& Spiga}]{Leconte2013}
Leconte, J., Forget, F., Charnay, B., {et~al.} 2013, Astronomy and Astrophysics, 554, \dodoi{10.1051/0004-6361/201321042}

\bibitem[{Lee {et~al.}(2011)Lee, Carslaw, Pringle, Mann, \& Spracklen}]{Lee2011}
Lee, L.~A., Carslaw, K.~S., Pringle, K.~J., Mann, G.~W., \& Spracklen, D.~V. 2011, Atmospheric Chemistry and Physics, 11, 12253, \dodoi{10.5194/acp-11-12253-2011}

\bibitem[{Lenardic {et~al.}(2016{\natexlab{a}})Lenardic, Crowley, Jellinek, \& Weller}]{Lenardic2016a}
Lenardic, A., Crowley, J.~W., Jellinek, A.~M., \& Weller, M. 2016{\natexlab{a}}, Astrobiology, 16, 551, \dodoi{10.1089/ast.2015.1378}

\bibitem[{Lenardic {et~al.}(2016{\natexlab{b}})Lenardic, Jellinek, Foley, O'Neill, \& Moore}]{Lenardic2016b}
Lenardic, A., Jellinek, A.~M., Foley, B., O'Neill, C., \& Moore, W.~B. 2016{\natexlab{b}}, Journal of Geophysical Research: Planets, 121, 1831, \dodoi{10.1002/2016JE005089}

\bibitem[{Lobo {et~al.}(2023)Lobo, Shields, Palubski, \& Wolf}]{Lobo2023}
Lobo, A.~H., Shields, A.~L., Palubski, I.~Z., \& Wolf, E. 2023, The Astrophysical Journal, 945, 161, \dodoi{10.3847/1538-4357/aca970}

\bibitem[{Martinez {et~al.}(2013)Martinez, Collette, Baudin, \& Christopoulou}]{pyDOE}
Martinez, J.-M., Collette, Y., Baudin, M., \& Christopoulou, M. 2013, {pyDOE}.
\newblock \url{https://pythonhosted.org/pyDOE/}

\bibitem[{McKinney(2010)}]{mck10}
McKinney, W. 2010, in Proceedings of the 9th {Python} in {Science} {Conference}, ed. S.~van~der Walt \& J.~Millman, 51--56

\bibitem[{Milanković(1941)}]{Milankovitch1941}
Milanković, M. 1941, Royal Serbian Academy Special Publication, 133, 1.
\newblock \url{https://cir.nii.ac.jp/crid/1572824499597240704}

\bibitem[{Mollière {et~al.}(2022)Mollière, Molyarova, Bitsch, Henning, Schneider, Kreidberg, Eistrup, Burn, Nasedkin, Semenov, Mordasini, Schlecker, Schwarz, Lacour, Nowak, \& Schulik}]{Molliere2022}
Mollière, P., Molyarova, T., Bitsch, B., {et~al.} 2022, The Astrophysical Journal, 934, 74, \dodoi{10.3847/1538-4357/ac6a56}

\bibitem[{Olson \& Christensen(2006)}]{Olson2006}
Olson, P., \& Christensen, U.~R. 2006, Earth and Planetary Science Letters, 250, 561, \dodoi{10.1016/j.epsl.2006.08.008}

\bibitem[{O'Malley-James \& Kaltenegger(2018)}]{OMalley-James2018}
O'Malley-James, J.~T., \& Kaltenegger, L. 2018, Astrobiology, 18, 1123, \dodoi{10.1089/ast.2017.1798}

\bibitem[{O’Hagan(2006)}]{OHagan2006}
O’Hagan, A. 2006, Reliability Engineering \& System Safety, 91, 1290, \dodoi{10.1016/j.ress.2005.11.025}

\bibitem[{Passegger {et~al.}(2024)Passegger, Suárez~Mascareño, Allart, González~Hernández, Lovis, Lavie, Silva, Müller, Tabernero, Cristiani, Pepe, Rebolo, Santos, Adibekyan, Alibert, Allende~Prieto, Barros, Bouchy, Castro-González, D’Odorico, Dumusque, Di~Marcantonio, Ehrenreich, Figueira, Génova~Santos, Lo~Curto, Martins, Mehner, Micela, Molaro, Nari, Nunes, Pallé, Poretti, Rodrigues, Sousa, Sozzetti, Udry, \& Zapatero~Osorio}]{Passegger2024}
Passegger, V.~M., Suárez~Mascareño, A., Allart, R., {et~al.} 2024, Astronomy \& Astrophysics, 684, A22, \dodoi{10.1051/0004-6361/202348592}

\bibitem[{Pedregosa {et~al.}(2011)Pedregosa, Varoquaux, Gramfort, Michel, Thirion, Grisel, Blondel, Prettenhofer, Weiss, Dubourg, Vanderplas, Passos, Cournapeau, Brucher, Perrot, \& Duchesnay}]{scikit-learn}
Pedregosa, F., Varoquaux, G., Gramfort, A., {et~al.} 2011, Journal of Machine Learning Research, 12, 2825

\bibitem[{Ramirez(2018)}]{Ramirez2018}
Ramirez, R. 2018, Geosciences, 8, 280, \dodoi{10.3390/geosciences8080280}

\bibitem[{Ramirez \& Kaltenegger(2014)}]{Ramirez2014}
Ramirez, R.~M., \& Kaltenegger, L. 2014, The Astrophysical Journal, 797, L25, \dodoi{10.1088/2041-8205/797/2/L25}

\bibitem[{Rasmussen \& Williams(2006)}]{books/lib/RasmussenW06}
Rasmussen, C.~E., \& Williams, C. K.~I. 2006, Gaussian {Processes} for {Machine} {Learning} (MIT Press).
\newblock \url{https://ui.adsabs.harvard.edu/abs/2006gpml.book.....R}

\bibitem[{Rugenstein {et~al.}(2020)Rugenstein, Bloch‐Johnson, Gregory, Andrews, Mauritsen, Li, Frölicher, Paynter, Danabasoglu, Yang, Dufresne, Cao, Schmidt, Abe‐Ouchi, Geoffroy, \& Knutti}]{Rugenstein2020}
Rugenstein, M., Bloch‐Johnson, J., Gregory, J., {et~al.} 2020, Geophysical Research Letters, 47, e2019GL083898, \dodoi{10.1029/2019GL083898}

\bibitem[{Schwieterman {et~al.}(2018)Schwieterman, Kiang, Parenteau, Harman, DasSarma, Fisher, Arney, Hartnett, Reinhard, Olson, Meadows, Cockell, Walker, Grenfell, Hegde, Rugheimer, Hu, \& Lyons}]{Schwieterman2018}
Schwieterman, E.~W., Kiang, N.~Y., Parenteau, M.~N., {et~al.} 2018, Astrobiology, 18, 663, \dodoi{10.1089/ast.2017.1729}

\bibitem[{Seager {et~al.}(2005)Seager, Turner, Schafer, \& Ford}]{Seager2005}
Seager, S., Turner, E., Schafer, J., \& Ford, E. 2005, Astrobiology, 5, 372, \dodoi{10.1089/ast.2005.5.372}

\bibitem[{Sexton {et~al.}(2019)Sexton, Karmalkar, Murphy, Williams, Boutle, Morcrette, Stirling, \& Vosper}]{Sexton2019}
Sexton, D. M.~H., Karmalkar, A.~V., Murphy, J.~M., {et~al.} 2019, Climate Dynamics, 53, 989, \dodoi{10.1007/s00382-019-04625-3}

\bibitem[{Shields {et~al.}(2014)Shields, Bitz, Meadows, Joshi, \& Robinson}]{shi14}
Shields, A., Bitz, C., Meadows, V., Joshi, M., \& Robinson, T. 2014, {\textbackslash}apjl, 785, L9, \dodoi{10.1088/2041-8205/785/1/L9}

\bibitem[{Spiegel {et~al.}(2010)Spiegel, Raymond, Dressing, Scharf, \& Mitchell}]{spi10}
Spiegel, D., Raymond, S., Dressing, C., Scharf, C., \& Mitchell, J. 2010, {\textbackslash}apj, 721, 1308, \dodoi{10.1088/0004-637X/721/2/1308}

\bibitem[{Spiegel {et~al.}(2008)Spiegel, Menou, \& Scharf}]{Spiegel2008}
Spiegel, D.~S., Menou, K., \& Scharf, C.~A. 2008, The Astrophysical Journal, 681, 1609, \dodoi{10.1086/588089}

\bibitem[{Spinelli {et~al.}(2023)Spinelli, Gallo, Haardt, Caldiroli, Biassoni, Borsa, \& Rauscher}]{Spinelli2023}
Spinelli, R., Gallo, E., Haardt, F., {et~al.} 2023, The Astronomical Journal, 165, 200, \dodoi{10.3847/1538-3881/acc336}

\bibitem[{Thomson \& Vallis(2019)}]{Thomson2019}
Thomson, S.~I., \& Vallis, G.~K. 2019, Quarterly Journal of the Royal Meteorological Society, 145, 2627, \dodoi{10.1002/qj.3582}

\bibitem[{Uusitalo {et~al.}(2015)Uusitalo, Lehikoinen, Helle, \& Myrberg}]{Uusitalo2015}
Uusitalo, L., Lehikoinen, A., Helle, I., \& Myrberg, K. 2015, Environmental Modelling and Software, 63, 24, \dodoi{10.1016/j.envsoft.2014.09.017}

\bibitem[{van~der Walt {et~al.}(2011)van~der Walt, Colbert, \& Varoquaux}]{van11}
van~der Walt, S., Colbert, S.~C., \& Varoquaux, G. 2011, Computing in Science Engineering, 13, 22, \dodoi{10.1109/MCSE.2011.37}

\bibitem[{Van~Eylen \& Albrecht(2015)}]{VanEylen2015}
Van~Eylen, V., \& Albrecht, S. 2015, The Astrophysical Journal, 808, 126, \dodoi{10.1088/0004-637X/808/2/126}

\bibitem[{Van~Eylen {et~al.}(2021)Van~Eylen, Astudillo-Defru, Bonfils, Livingston, Hirano, Luque, Lam, Justesen, Winn, Gandolfi, Nowak, Palle, Albrecht, Dai, Estrada, Owen, Foreman-Mackey, Fridlund, Korth, Mathur, Forveille, Mikal-Evans, Osborne, Ho, Almenara, Artigau, Barragán, Barros, Bouchy, Cabrera, Caldwell, Charbonneau, Chaturvedi, Cochran, Csizmadia, Damasso, Delfosse, De~Medeiros, Díaz, Doyon, Esposito, Fűrész, Figueira, Georgieva, Goffo, Grziwa, Guenther, Hatzes, Jenkins, Kabath, Knudstrup, Latham, Lavie, Lovis, Mennickent, Mullally, Murgas, Narita, Pepe, Persson, Redfield, Ricker, Santos, Seager, Serrano, Smith, Mascareño, Subjak, Twicken, Udry, Vanderspek, \& Osorio}]{VanEylen2021}
Van~Eylen, V., Astudillo-Defru, N., Bonfils, X., {et~al.} 2021, Monthly Notices of the Royal Astronomical Society, stab2143, \dodoi{10.1093/mnras/stab2143}

\bibitem[{Vervoort {et~al.}(2022)Vervoort, Horner, Kane, Kirtland~Turner, \& Gilmore}]{Vervoort2022}
Vervoort, P., Horner, J., Kane, S.~R., Kirtland~Turner, S., \& Gilmore, J.~B. 2022, The Astronomical Journal, 164, 130, \dodoi{10.3847/1538-3881/ac87fd}

\bibitem[{Way \& Georgakarakos(2016)}]{Way2016}
Way, M.~J., \& Georgakarakos, N. 2016, The Astrophysical Journal Letters, 835, 1, \dodoi{10.3847/2041-8213/835/1/L1}

\bibitem[{Way {et~al.}(2023)Way, Georgakarakos, \& Clune}]{Way2023}
Way, M.~J., Georgakarakos, N., \& Clune, T.~L. 2023, The Astronomical Journal, 166, 227, \dodoi{10.3847/1538-3881/ad0373}

\bibitem[{Way {et~al.}(2017)Way, Aleinov, Amundsen, Chandler, Clune, Genio, Fujii, Kelley, Kiang, Sohl, \& Tsigaridis}]{Way2017}
Way, M.~J., Aleinov, I., Amundsen, D.~S., {et~al.} 2017, The Astrophysical Journal Supplement Series, 231, 12, \dodoi{10.3847/1538-4365/aa7a06}

\bibitem[{Willmott \& Matsuura(2018)}]{Willmott2018}
Willmott, C.~J., \& Matsuura, K. 2018, Terrestrial {Precipitation}: 1900-2017 {Gridded} {Monthly} {Time} {Series}.
\newblock \url{http://climate.geog.udel.edu/html_pages/download.html}

\bibitem[{Wolf {et~al.}(2017)Wolf, Shields, Kopparapu, Haqq-Misra, \& Toon}]{Wolf2017}
Wolf, E.~T., Shields, A.~L., Kopparapu, R.~K., Haqq-Misra, J., \& Toon, O.~B. 2017, The Astrophysical Journal, 837, 107, \dodoi{10.3847/1538-4357/aa5ffc}

\end{thebibliography}
\bibliographystyle{aasjournal}



\end{CJK*}
\end{document}